\documentclass[11pt,a4paper]{article}
\usepackage{jcappub}

\usepackage{amstext}
\usepackage{amssymb}
\usepackage{graphicx}
\usepackage{latexsym}
\usepackage{mathrsfs}
\usepackage{calrsfs}
\usepackage{amsfonts}

{\count255=\time\divide\count255 by 60 \xdef\hourmin{\number\count255}
  \multiply\count255 by-60\advance\count255 by\time
  \xdef\hourmin{\hourmin:\ifnum\count255<10 0\fi\the\count255}}

\def\rd{ {\rm d}}
\def\vev#1{ \left\langle #1 \right \rangle }

\def\tobs{\Delta}

\begin{document}

\title{The Topology and Size of the Universe from CMB Temperature and Polarization Data}

\author[a]{Grigor Aslanyan,}
\author[b]{Aneesh V.~Manohar,}
\author[b]{Amit P.S. Yadav}

\affiliation[a]{Department of Physics, University of Auckland, Private Bag 92019, Auckland, New Zealand}
\affiliation[b]{Department of Physics, University of California at San Diego,
  La Jolla, CA 92093\vspace{4pt} }
  
\emailAdd{g.aslanyan@auckland.ac.nz}
\emailAdd{amanohar@ucsd.edu}
\emailAdd{ayadav@ucsd.edu}

\abstract{We analyze seven year and nine year WMAP temperature maps for signatures of three finite flat topologies $\mathcal{M}_0=\mathbb{T}^3$, $\mathcal{M}_1=\mathbb{T}^2\times\mathbb{R}^1$, and $\mathcal{M}_2=S^1\times\mathbb{R}^2$. We use Monte-Carlo simulations with the Feldman-Cousins method to obtain confidence intervals for the size of the topologies considered. We analyze the V, W, and Q frequency bands along with the ILC map and find no significant difference in the results. The $95.5\%$ confidence level lower bound on the size of the topology is $1.5L_0$ for $\mathcal{M}_0$, $1.4L_0$ for $\mathcal{M}_1$, and $1.1L_0$ for $\mathcal{M}_2$, where $L_0$ is the radius of the last scattering surface. Our results agree very well with the recently released results from the Planck temperature data. We show that the likelihood function is not Gaussian in the size, and therefore simulations are important for obtaining accurate bounds on the size. We then introduce the formalism for including polarization data in the analysis. The improvement that we find from WMAP polarization maps is small because of the high level of instrumental noise, but our forecast for Planck maps shows a much better improvement on the lower bound for $L$. For the $\mathcal{M}_0$ topology we expect an improvement on the lower bound of $L$ from $1.7L_0$ to $1.9L_0$ at $95.5\%$ confidence level. Using both polarization and temperature data is important because it tests the hypothesis that deviations in the $TT$ spectrum at small $l$ originate in the primordial perturbation spectrum.
}

\arxivnumber{1304.1811}

\maketitle

\section{Introduction}

The global topology of the universe has been extensively studied in the literature since precise measurements of the cosmic microwave background (CMB) radiation became available from the COBE and WMAP satellites \cite{deOlivieraCosta:1994eb,deOliveiraCosta:1995td,Phillips:2004nc,Kunz:2005wh,Aurich:2007yx,Aslanyan:2011zp,Roukema:2010mw,Luminet:2003dx,Caillerie:2007gd,Lew:2008yz,Aurich:2004fq,Roukema:2011xm,2011MNRAS.tmp..137B, Aurich:2013}. The latest temperature data from the Planck satellite has also been used to constrain the topology of the universe \cite{planck_topology}. There are two main methods for this study: the circles in the sky method and low-$l$ correlations of CMB anisotropies. The first method is based on the fact that if the global structure of the universe is smaller than the last scattering surface (LSS); the LSS intersects itself resulting in matching circles in different directions on the sky. This method has been used to put bounds on the size of the universe with the most recent bound of $27.9\,\text{Gpc}$ from the WMAP seven year data \cite{2011MNRAS.tmp..137B}. The authors of \cite{Bielewicz:2011jz} have discussed the usage of the method for CMB polarization data. This method, however, does not work if the global topology of the universe is bigger than the diameter of the LSS. The radius of the LSS is $L_0=14.4\,\text{Gpc}$, so the current bound from this method is already very close to the maximum value of $28.8\,\text{Gpc}$ possibly detectable by the method.

The second main method to detect the topology of the universe uses the correlations of the CMB anisotropies on large scales. This method has been discussed in detail in \cite{Aslanyan:2011zp}, where it was also used to analyze three different flat topologies of the universe: $\mathcal{M}_0=\mathbb{T}^3$, $\mathcal{M}_1=\mathbb{T}^2\times\mathbb{R}^1$, and $\mathcal{M}_2=S^1\times\mathbb{R}^2$. The topology $\mathcal{M}_i$ has $i$ infinite dimensions with the other $3-i$ dimensions compactified to the same size $L$. Ref.~\cite{Aslanyan:2011zp} found an improvement of about $20$ in $-2\ln\mathcal{L}$ for the $\mathcal{M}_1$ topology with size $1.9L_0$ compared to the infinite universe. It is therefore of importance to perform other independent tests of these results.

Previous analyses of the topology of the universe have used only the temperature data from CMB experiments, the reason being that the currently available polarization data is very noisy. However,  polarization data from Planck will have much less instrumental noise than  currently available data from WMAP, and  can be used to improve the constraints on the topology of the universe. Signals for non-trivial topology can be interpreted, more generally, as evidence that the primordial fluctuation spectrum deviates from the standard prediction of inflationary cosmology at small values of $l$. A particular scenario, such as the torus topology we consider, provides a specific  model with a few parameters that can be fit to observations. Polarization provides a very important check that  anomalies in the $TT$ spectrum 
are primordial, since in this case, deviations in the $TT$, $TE$ and $EE$ spectrum are related as they have a common origin.
This would not be the case if deviations in the $TT$ spectrum arose from late-time effects after decoupling.

This paper serves two purposes. Firstly, we develop the formalism that can be used to analyze the polarization data in addition to the temperature data to improve the constraints on the topology of the universe. Although this may not give a big improvement in the results with the currently available polarization data from WMAP, it will be of invaluable use with the Planck polarization data that is expected to be released soon. Secondly, we use the Feldman-Cousins method \cite{Feldman:1997qc} to accurately estimate the confidence intervals from the likelihood function using simulated sky maps. The previous analysis \cite{Aslanyan:2011zp} used the ILC temperature map only from seven year WMAP data. We repeat this analysis with a slight improvement in the technique using the V, W, and Q frequency bands as well to test if the observed effect is based on some anomaly in the ILC map. We also analyze the nine year release of these maps to check if there is any improvement in the results from new temperature maps. We then analyze the WMAP data using the polarization maps as well, and we give a forecast for the Planck data.

The paper is organized as follows. In Section \ref{likelihood_sec} we describe the likelihood calculation for temperature and polarization maps, in Section \ref{data_analysis_sec} we discuss the details of the data analysis, in Section \ref{simulations_sec} we describe the usage of the Feldman-Cousins method and how to simulate finite sky maps. Our results for WMAP data are summarized in Section \ref{results_sec}. We then compare our results with the most recent results from Planck temperature data and give a forecast for Planck polarization data in Section \ref{forecast_sec}, and we conclude in Section \ref{conclusions_sec}.

\section{Likelihood Calculation for Temperature and Polarization Maps}\label{likelihood_sec}

The likelihood calculation for finite topologies using the temperature data only is described in detail in \cite{Aslanyan:2011zp}. Here we generalize this analysis to include the polarization data as well. We work with the torus topology $\mathbb{T}^3$, with sides $L_1, L_2, $ and $L_3$. The other cases can be obtained from this by taking some $L_i \to \infty.$

The temperature map $T$, and polarization maps $Q$ and $U$ are measured by a CMB experiment. These maps can be decomposed into spherical harmonics
\begin{equation}
T(\hat{\mathbf{n}})=\sum_{lm}T_{lm}Y_{lm}(\hat{\mathbf{n}})\,,
\end{equation}
\begin{equation}
Q(\hat{\mathbf{n}})\pm iU(\hat{\mathbf{n}})=\sum_{lm}{}_{\mp2}a_{lm}\;{}_{\mp2}Y_{lm}(\hat{\mathbf{n}})
\end{equation}
where ${}_sY_{lm}$ are the spin-weighted spherical harmonics. The polarization coefficients can be further decomposed into real and imaginary parts ($E$ and $B$ modes) \cite{Page:2006hz}
\begin{equation}
{}_{\pm2}a_{lm}=E_{lm}\pm iB_{lm}
\end{equation}
which results in
\begin{equation}
Q(\hat{\mathbf{n}})=\frac{1}{2}\sum_{lm}\left[E_{lm}({}_{+2}Y_{lm}(\hat{\mathbf{n}})+{}_{-2}Y_{lm}(\hat{\mathbf{n}}))+iB_{lm}({}_{+2}Y_{lm}(\hat{\mathbf{n}})-{}_{-2}Y_{lm}(\hat{\mathbf{n}}))\right]\,,
\end{equation}
\begin{equation}
U(\hat{\mathbf{n}})=\frac{i}{2}\sum_{lm}\left[E_{lm}({}_{+2}Y_{lm}(\hat{\mathbf{n}})-{}_{-2}Y_{lm}(\hat{\mathbf{n}}))+iB_{lm}({}_{+2}Y_{lm}(\hat{\mathbf{n}})+{}_{-2}Y_{lm}(\hat{\mathbf{n}}))\right]\,.
\end{equation}

All of these maps are, in general, correlated with each other. In order to calculate the likelihood $\mathcal{L}(\mathbf{m}|S)\rd\mathbf{m}$, one needs to construct a big covariance matrix (including noise) that will include the $T$, $Q$, and $U$ modes
\begin{equation}
\mathcal{L}(\mathbf{m}|S)\rd\mathbf{m}=\frac{\exp\left[-\frac{1}{2}\mathbf{m}^t(S+N)^{-1}\mathbf{m}\right]}{(2\pi)^{3n_p/2}|S+N|^{1/2}}\rd\mathbf{m}
\end{equation}
where $\mathbf{m}=(\mathbf{T},\mathbf{Q},\mathbf{U})$, $S$ and $N$ are the signal and noise covariance matrices, respectively, $n_p$ is the number of pixels, and symbol $|..|$ stands for determinant. The signal covariance matrix is calculated more easily in harmonic space. Let us denote
\begin{equation}\label{cov_mat_def}
\vev{\mathcal{X}_{lm}\mathcal{Y}_{l^\prime m^\prime}^*}=M^{\mathcal{XY}}_{lml^\prime m^\prime}
\end{equation}
where $\mathcal{X}$ and $\mathcal{Y}$ denote $T$, $E$, and $B$. For an infinite universe rotational invariance ensures that these matrices are diagonal in harmonic $(\ell,m)$ space, 
\begin{equation}
M^{\mathcal{XY}}_{lml^\prime m^\prime}=\delta_{ll^\prime}\delta_{mm^\prime}C^{\mathcal{XY}}_{l}
\end{equation}
while for the finite flat topologies the isotropy of space is broken and these matrices acquire non-zero off-diagonal elements \cite{Aslanyan:2011zp}. The temperature-temperature correlation matrix has been derived in \cite{Aslanyan:2011zp}. This result can be easily generalized to include the $E$ and $B$ modes as well
\begin{eqnarray}\label{M_torus}
&& M^{\mathcal{XY}}_{lml'm'} = (4\pi)^2(-i)^li^{l^\prime}  \frac{1}{L_1L_2L_3}\sum_{\mathbf{k}}P_\zeta(k)g^\mathcal{X}_l(k)g^{\mathcal{Y}*}_{l^\prime}(k)
 Y_{lm}^*(\hat{\mathbf{k}})Y_{l^\prime m^\prime}(\hat{\mathbf{k}})\,.
\end{eqnarray}
where $L_1$, $L_2$, $L_3$ are the sizes of the three sides of the torus topology, $P_\zeta$ is the primordial power spectrum of the gauge invariant curvature perturbations $\zeta$, $g^\mathcal{X}_l$ and $g^\mathcal{Y}_l$ are the radiative transfer functions for modes $\mathcal{X}$ and $\mathcal{Y}$. The sum over $\mathbf{k}=(k_1,k_2,k_3)$ runs over the values
\begin{equation}\label{k_spectrum}
k_1=\frac{2\pi}{L_1}n_1,\quad k_2=\frac{2\pi}{L_2}n_2,\quad k_3=\frac{2\pi}{L_3}n_3
\end{equation}
where $n_1$, $n_2$, $n_3$ are integers.

All of the symmetry arguments described in \cite{Aslanyan:2011zp} for $M^{TT}$ can be generalized to the other modes. In particular, all of these matrices are real, and $M^{TT}$, $M^{EE}$, and $M^{BB}$ are symmetric. From now on we will assume no parity violation, which implies $M^{TB}=0$, $M^{EB}=0.$\footnote{CMB lensing introduces non-zero contributions into these matrices, however, lensing effects are negligible on the large scales we consider.} No gravitational waves have been detected so far by CMB measurements \cite{2012arXiv1212.5226H} so we will assume that $M^{BB}=0$ as well.

For low-$l$ modes, the noise in the temperature map can be ignored, while there is a significant noise in polarization maps. For this reason it is easier to calculate the likelihood by decomposing polarization into correlated and uncorrelated parts with temperature \cite{Page:2006hz}. Let us define
\begin{equation}
\tilde{E}_{lm}=E_{lm}-M^{TE}_{l^\prime m^\prime lm}(M^{TT}_{l^\prime m^\prime l^{\prime\prime}m^{\prime\prime}})^{-1}T_{l^{\prime\prime}m^{\prime\prime}}
\end{equation}
which gives
\begin{equation}
\vev{\tilde{E}_{lm}T_{l^\prime m^\prime}^*}=0\,,
\end{equation}
\begin{equation}\label{new_EE}
\vev{\tilde{E}_{lm}\tilde{E}_{l^\prime m^\prime}^*}\equiv M^{\tilde{E}\tilde{E}}_{lml^\prime m^\prime}=M^{EE}_{lml^\prime m^\prime}-M^{TE}_{l^{\prime\prime}m^{\prime\prime}lm}(M^{TT}_{l^{\prime\prime}m^{\prime\prime}l^{\prime\prime\prime}m^{\prime\prime\prime}})^{-1}M^{TE}_{l^{\prime\prime\prime}m^{\prime\prime\prime}l^\prime m^\prime}\,.
\end{equation}

We now define
\begin{equation}
\tilde{Q}(\hat{\mathbf{n}})=\frac{1}{2}\sum_{lm}\left[\tilde{E}_{lm}({}_{+2}Y_{lm}(\hat{\mathbf{n}})+{}_{-2}Y_{lm}(\hat{\mathbf{n}}))+iB_{lm}({}_{+2}Y_{lm}(\hat{\mathbf{n}})-{}_{-2}Y_{lm}(\hat{\mathbf{n}}))\right]\,,
\end{equation}
\begin{equation}
\tilde{U}(\hat{\mathbf{n}})=\frac{i}{2}\sum_{lm}\left[\tilde{E}_{lm}({}_{+2}Y_{lm}(\hat{\mathbf{n}})-{}_{-2}Y_{lm}(\hat{\mathbf{n}}))+iB_{lm}({}_{+2}Y_{lm}(\hat{\mathbf{n}})+{}_{-2}Y_{lm}(\hat{\mathbf{n}}))\right]\,.
\end{equation}
Then in pixel space we get, $ \vev{\tilde{Q}(\hat{\mathbf{n}}_i)T(\hat{\mathbf{n}}_j)}=0\,,
\vev{\tilde{U}(\hat{\mathbf{n}}_i)T(\hat{\mathbf{n}}_j)}=0\,.$
This allows for the likelihood function to be decomposed into a product of two factors
\begin{equation}
\mathcal{L}(\mathbf{m}|S)\rd\mathbf{m}=\frac{\exp\left[-\frac{1}{2}\tilde{\mathbf{m}}^t(\tilde{S}_P+N_P)^{-1}\tilde{\mathbf{m}}\right]}{(2\pi)^{n_p}|\tilde{S}_P+N_P|^{1/2}}\rd\tilde{\mathbf{m}}\,\frac{\exp\left[-\frac{1}{2}\mathbf{T}^tS_T^{-1}\mathbf{T}\right]}{(2\pi)^{n_p/2}|S_T|^{1/2}}\rd\mathbf{T}
\end{equation}
where $\tilde{\mathbf{m}}=(\tilde{\mathbf{Q}},\tilde{\mathbf{U}})$. We have ignored the noise in temperature. The noise matrix for the new $\tilde{Q}$ and $\tilde{U}$ variables is the same as for the original $Q$ and $U$. The new signal covariance matrix can be calculated as follows
\begin{eqnarray}
\vev{\tilde{Q}(\hat{\mathbf{n}}_i)\tilde{Q}(\hat{\mathbf{n}}_j)}=&&\frac{1}{4}\sum_{lml^\prime m^\prime}\left[M^{\tilde{E}\tilde{E}}_{lml^\prime m^\prime}({}_{+2}Y_{lm}(\hat{\mathbf{n}}_i)+{}_{-2}Y_{lm}(\hat{\mathbf{n}}_i))({}_{+2}Y_{l^\prime m^\prime}^*(\hat{\mathbf{n}}_j)+{}_{-2}Y_{l^\prime m^\prime}^*(\hat{\mathbf{n}}_j))\right. \nonumber \\
&&\left.-M^{BB}_{lml^\prime m^\prime}({}_{+2}Y_{lm}(\hat{\mathbf{n}}_i)-{}_{-2}Y_{lm}(\hat{\mathbf{n}}_i))({}_{+2}Y_{l^\prime m^\prime}^*(\hat{\mathbf{n}}_j)-{}_{-2}Y_{l^\prime m^\prime}^*(\hat{\mathbf{n}}_j))\right]\,,
\end{eqnarray}
\begin{eqnarray}
\vev{\tilde{Q}(\hat{\mathbf{n}}_i)\tilde{U}(\hat{\mathbf{n}}_j)}=&&\frac{i}{4}\sum_{lml^\prime m^\prime}\left[M^{\tilde{E}\tilde{E}}_{lml^\prime m^\prime}({}_{+2}Y_{lm}(\hat{\mathbf{n}}_i)+{}_{-2}Y_{lm}(\hat{\mathbf{n}}_i))({}_{+2}Y_{l^\prime m^\prime}^*(\hat{\mathbf{n}}_j)-{}_{-2}Y_{l^\prime m^\prime}^*(\hat{\mathbf{n}}_j))\right. \nonumber \\
&&\left.-M^{BB}_{lml^\prime m^\prime}({}_{+2}Y_{lm}(\hat{\mathbf{n}}_i)-{}_{-2}Y_{lm}(\hat{\mathbf{n}}_i))({}_{+2}Y_{l^\prime m^\prime}^*(\hat{\mathbf{n}}_j)+{}_{-2}Y_{l^\prime m^\prime}^*(\hat{\mathbf{n}}_j))\right]\,,
\end{eqnarray}
\begin{eqnarray}
\vev{\tilde{U}(\hat{\mathbf{n}}_i)\tilde{U}(\hat{\mathbf{n}}_j)}=&&-\frac{1}{4}\sum_{lml^\prime m^\prime}\left[M^{\tilde{E}\tilde{E}}_{lml^\prime m^\prime}({}_{+2}Y_{lm}(\hat{\mathbf{n}}_i)-{}_{-2}Y_{lm}(\hat{\mathbf{n}}_i))({}_{+2}Y_{l^\prime m^\prime}^*(\hat{\mathbf{n}}_j)-{}_{-2}Y_{l^\prime m^\prime}^*(\hat{\mathbf{n}}_j))\right. \nonumber \\
&&\left.-M^{BB}_{lml^\prime m^\prime}({}_{+2}Y_{lm}(\hat{\mathbf{n}}_i)+{}_{-2}Y_{lm}(\hat{\mathbf{n}}_i))({}_{+2}Y_{l^\prime m^\prime}^*(\hat{\mathbf{n}}_j)+{}_{-2}Y_{l^\prime m^\prime}^*(\hat{\mathbf{n}}_j))\right]
\end{eqnarray}
where $M^{\tilde{E}\tilde{E}}_{lml^\prime m^\prime}$ is given by (\ref{new_EE}).

Following \cite{Page:2006hz}, we further rewrite the polarization part of the likelihood as follows
\begin{equation}
\mathcal{L}(\tilde{\mathbf{m}}|\tilde{S}_P)\rd\tilde{\mathbf{m}}=\frac{\exp\left[-\frac{1}{2}(N_P^{-1}\tilde{\mathbf{m}})^t(N_P^{-1}\tilde{S}_PN_P^{-1}+N_P^{-1})^{-1}(N_P^{-1}\tilde{\mathbf{m}})\right]}{(2\pi)^{n_p}|N_P^{-1}\tilde{S}_PN_P^{-1}+N_P^{-1}|^{1/2}}|N_P|^{-1}\rd\tilde{\mathbf{m}}
\end{equation}
which is numerically more tractable since it contains only $N_P^{-1}$. The temperature part of the likelihood calculation is described in detail in \cite{Aslanyan:2011zp}.

Since the likelihood is eventually calculated in pixel space, no changes are required to the formalism for a masked sky, we simply need to keep only the unmasked pixels in temperature and polarization maps and the corresponding signal and noise matrices. Masks usually have a large effect on $B$ modes, such as leakage from $E$ to $B$ modes, however this is not relevant for our analysis since we are ignoring the $B$ modes altogether.

\section{Data Analysis}\label{data_analysis_sec}

We calculate the radiative transfer functions in (\ref{M_torus}) using the CAMB software \cite{Lewis:1999bs}. We calculate the likelihood function using a modification of the WMAP likelihood code \cite{Jarosik:2010iu,Larson:2010gs,Komatsu:2010fb} to include the off-diagonal elements of the covariance matrices in harmonic space. The rest of the code is unaffected. Since our analysis is sensitive to large scales only, we use the low-$l$ part of the code which calculates the likelihood in pixel space. The temperature map used is the Internal Linear Combination (ILC) map smoothed to $9.183^\circ$ and degraded to $N_{\rm side}=16$ in HEALPIX format \cite{Gorski:2004by}. To check the consistency between different frequency bands we also do the analysis on V, W, and Q maps separately, smoothed and degraded in the same way. The temperature map is masked with the Kp2 mask, after which $2482$ pixels are left. The noise in the temperature map at that low resolution is negligible, however a $1\,\mu K$ white noise is added to each pixel (and a corresponding term to the covariance matrix) to aid the numerical regularization of the matrix inversion \cite{Hinshaw:2006ia}. As discussed in \cite{Aslanyan:2013zs}, the noise in the low resolution V, W, and Q maps is smaller than $1\,\mu K$ meaning that these maps can be analyzed in the same way as the ILC map.

The monopole and dipole terms are removed from the original full sky maps, however it is no longer true after applying the mask. These contributions are therefore marginalized over by  introducing large variance monopole and dipole terms into the covariance matrix. It is also necessary to marginalize over the residual foreground contamination. An extra parameter $\xi$ is introduced into the likelihood function
\begin{equation}\label{likelihood_fore}
\mathcal{L}(C, \xi)=\frac{1}{(2\pi)^{n_p/2}(\det C)^{1/2}}\exp\left(-\frac{1}{2}(\tobs-\xi\tobs_f)^T C^{-1}(\tobs-\xi\tobs_f)\right)
\end{equation}
where $\tobs$ is the data vector in pixel space, $\tobs_f$ is a foreground template, $C$ is the covariance matrix, $n_p$ is the number of pixels. We perform foreground marginalization by integrating the likelihood function (\ref{likelihood_fore}) over $\xi$. This can be done analytically with the result
\begin{equation}\label{likelihood_marginalized}
\mathcal{L}(C)=\frac{1}{(2\pi)^{n_p/2}(\det C)^{1/2}}\sqrt{\frac{2\pi}{\tobs_f^TC^{-1}\tobs_f}}\exp\left(-\frac{1}{2}\left(\tobs^TC^{-1}\tobs-\frac{(\tobs^TC^{-1}\tobs_f)^2}{\tobs_f^TC^{-1}\tobs_f}\right)\right)\,.
\end{equation}

We obtain the foreground template by taking the difference between the V band and the ILC map. This is the same template as the one used in WMAP low-$l$ likelihood code.

The polarization map used in the low resolution likelihood code is formed by a weighted combination of Ka, Q, and V bands, masked by the P06 mask and degraded to $N_{\rm side}=8$ \cite{Page:2006hz}. Terms up to $l_{\rm max}=30$ are included in the temperature analysis and $l_{\rm max}=23$ in the polarization analysis. It has been shown in \cite{Aslanyan:2011zp} that cutting off at $l=30$ has no significant effect on the temperature analysis and that terms up to $l=20$ already contain the essential effects of the topology analysis.

\begin{figure}
\centering
\includegraphics[width=7.8cm]{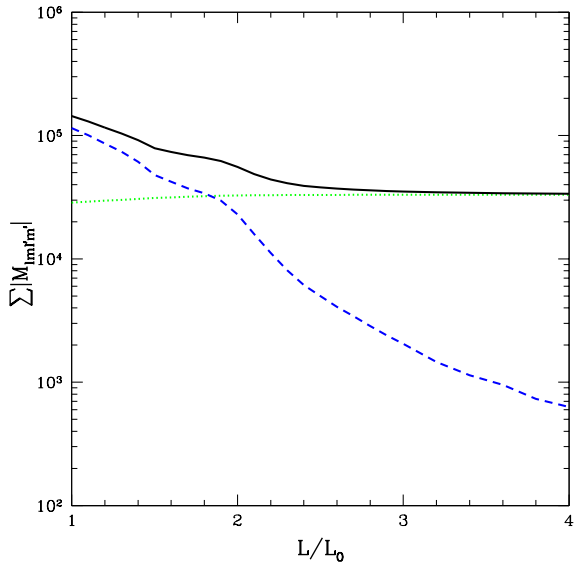}
\caption{\label{diagnostics_fig} The total power $\sum_{lml^\prime m^\prime} |M^T_{lml^\prime m^\prime}|$ as a function of $L/L_0$ for the $\mathcal{M}_0$ topology (the units are $\mu K^2$). The green dotted curve contains diagonal elements only $(l,m)=(l^\prime,m^\prime)$, the blue dashed curve contains off-diagonal elements $(l,m)\neq(l^\prime,m^\prime)$, the black solid curve contains all the elements.
}
\end{figure}

To estimate the importance of the off-diagonal terms of the covariance matrix $M_{lml^\prime m^\prime}$, in Fig. \ref{diagnostics_fig} we plot the total power $\sum_{lml^\prime m^\prime} |M^T_{lml^\prime m^\prime}|$ as a function of $L/L_0$ for the $\mathcal{M}_0$ topology, and we separate the diagonal terms from the off-diagonal ones (a similar test has been done in \cite{Phillips:2004nc} with similar results). As we can see, for sizes below $1.8L_0$ the off-diagonal terms dominate, for sizes up to $2L_0$ the off-diagonal terms are of similar magnitude as the diagonal ones, while for sizes above $2L_0$ the off-diagonal elements fall off quickly and rapidly become negligible. As can be seen later, our analysis is most sensitive to sizes below $2L_0$, so keeping the off-diagonal elements in the analysis is crucial.

\begin{figure}
\centering
\includegraphics[width=7.5cm]{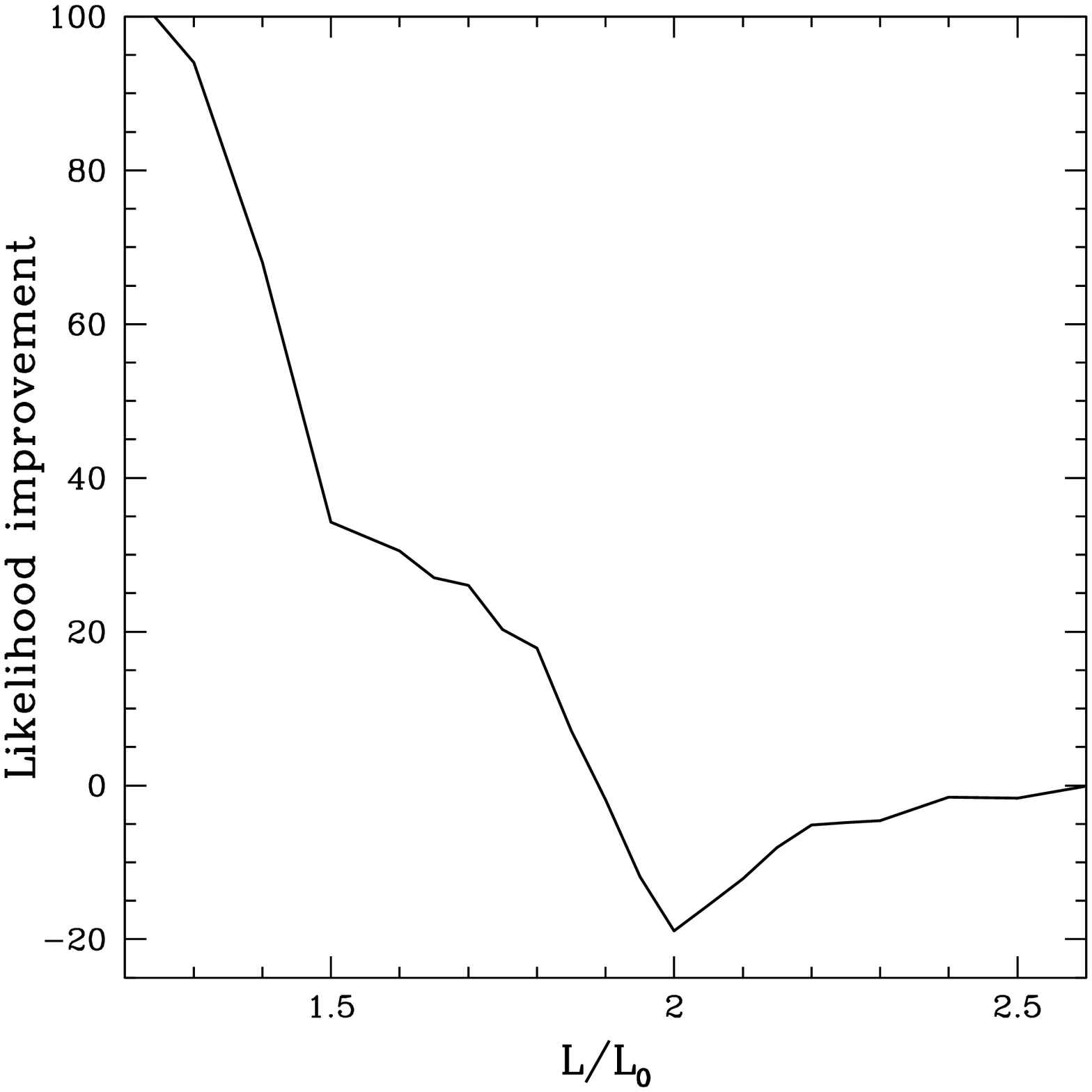}
\includegraphics[width=7.5cm]{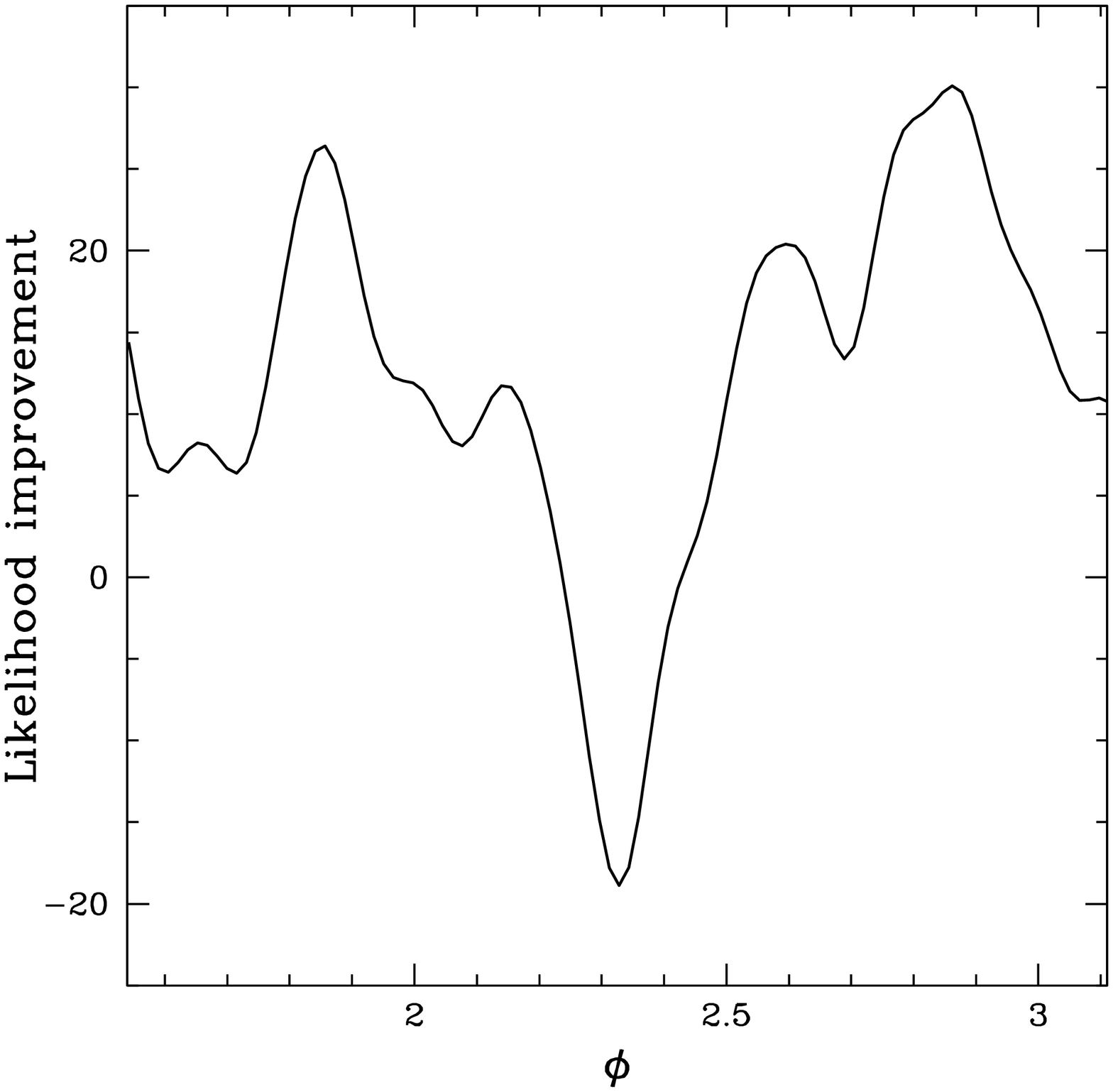}
\includegraphics[width=7.5cm]{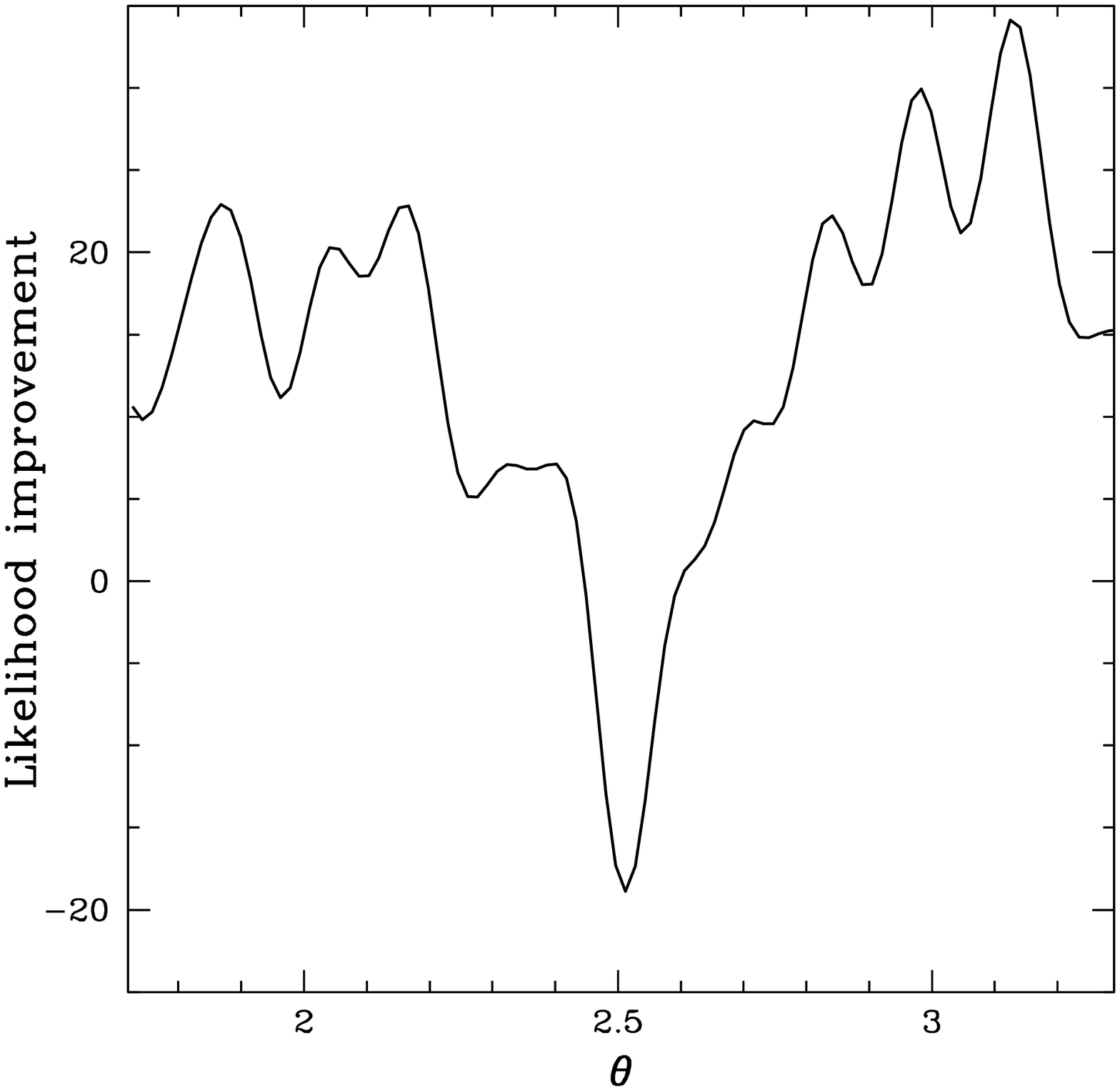}
\includegraphics[width=7.5cm]{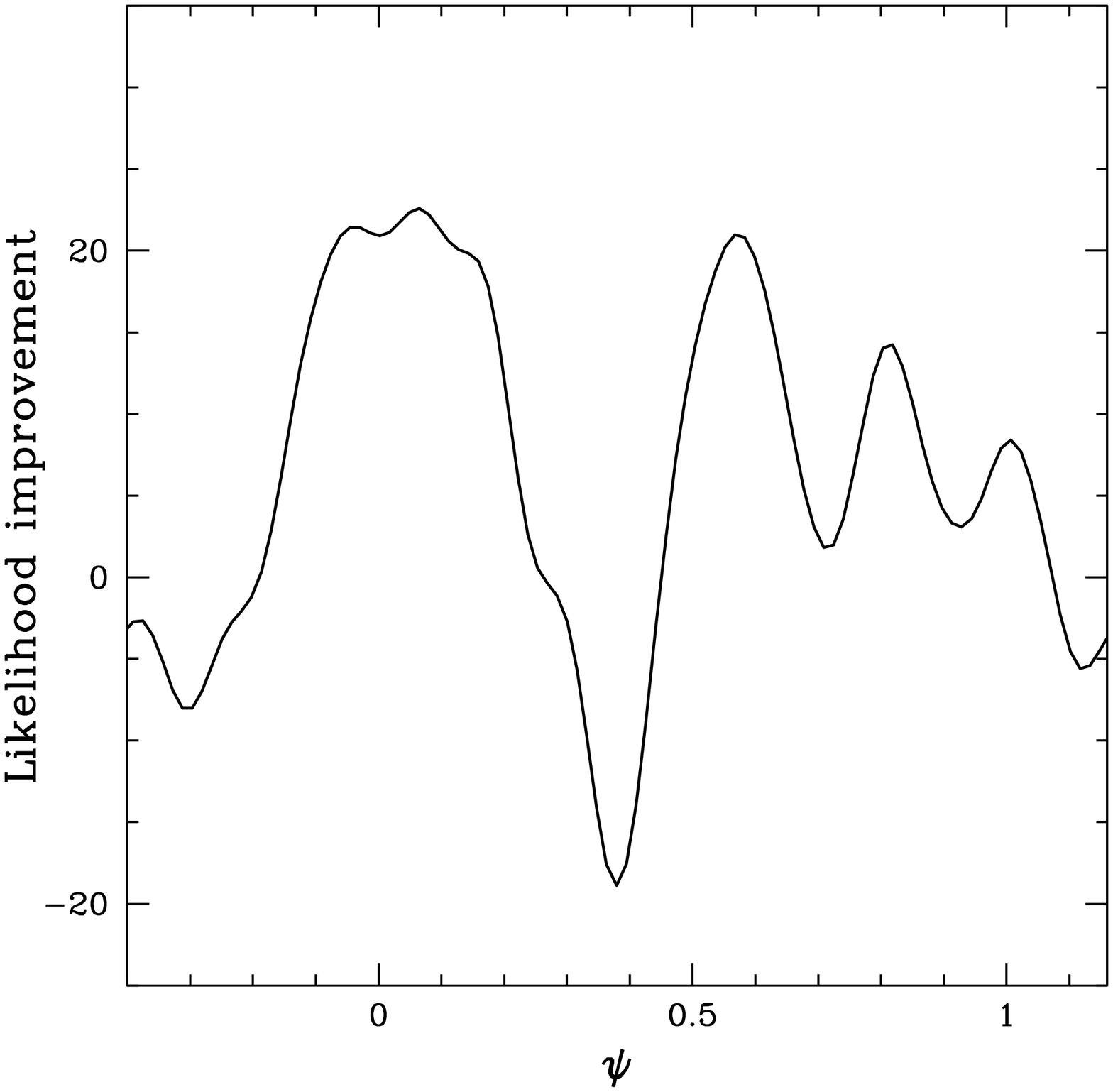}
\caption{\label{likelihood_variation_fig} Variation of $\Delta\mathcal{L}=-2\ln\mathcal{L}+2\ln\mathcal{L}_{\infty}$ as a function of the size (top left), Euler angle $\phi$ (top right), $\theta$ (bottom left), and $\psi$ (bottom right). For each plot all the other parameters are kept fixed. All of the parameters are varied around the best fit point for the seven year ILC map for the $\mathcal{M}_0$ topology. Likelihood includes temperature only.
}
\end{figure}

We do a scan over the sizes of the three topologies $\mathcal{M}_0$, $\mathcal{M}_1$, and $\mathcal{M}_2$. These topologies break the isotropy of space, therefore we need to also scan over different orientations of the topologies. We describe the orientation of the torus by three Euler angles $(\phi,\theta,\psi)$. The notation is the same as in \cite{Aslanyan:2011zp}. We denote the coordinate frame of the CMB data by $x$, $y$, $z$, and the coordinate frame of the torus by $x^\prime$, $y^\prime$, $z^\prime$ (these axes coincide with the axes of the torus. For $\mathcal{M}_1$ the $z^\prime$ axis points in the infinite direction, while for $\mathcal{M}_2$ it points in the finite direction.). We start by the two frames aligned with each other, then rotate the torus counterclockwise around the $z$-axis by angle $\phi$, then around the new $x$-axis by angle $\theta$, then around the new $z$-axis by angle $\psi$. Some of these angles are equivalent to each other because of the symmetries of the tori. For each topology we need to minimize $-2\ln\mathcal{L}$ over the size and direction. In Fig. \ref{likelihood_variation_fig} we show the dependence of the likelihood on all four new parameters separately. The scan is done for the $\mathcal{M}_0$ topology with the seven year ILC map. We have performed a fine scan around the best fit point (see Section \ref{results_sec}). As we can see, the likelihood oscillates very rapidly as a function of the Euler angles, therefore using a minimization algorithm on the whole set of directions may not find the global minimum (this issue has also been discussed in \cite{Aslanyan:2011zp}). For that reason the minimization is done in two steps. We first construct a grid in the Euler angle space with step size of $\pi/20$ and identify the points in the grid that correspond to equivalent directions of the torus. The step size is chosen to be approximately equal to half of the oscillation length, so that each grid cell contains not more than one local minimum. We then take the point at which the minimum is reached after the initial scan, and all of the points that are above the minimum by not more than $5$, and do a second scan around these points now using the CERN MINUIT package for function minimization \cite{minuit}.\footnote{By doing a second scan over more points of the initial grid we have verified that our scan strategy is good enough to find the global minimum.} The dependence of the likelihood on the size $L$ is not as strong, so we do a scan over the sizes with a step $0.1L_0$.

\section{Feldman-Cousins Method and Simulations}\label{simulations_sec}

Once the likelihood function is calculated and maximized on the whole parameter space, we need to derive the confidence intervals for the size of the topologies analyzed. The likelihood depends on the size of the topology in a very complicated way. Namely, the likelihood depends on the covariance matrix in pixel space, which depends on the covariance matrix in harmonic space, the dependence of which on the size of the topology is given by eq. (\ref{M_torus}). The previous analysis \cite{Aslanyan:2011zp} used the maximum likelihood method relying on the assumption that the likelihood ratios have a gaussian distribution in terms of the size. Although this assumption is valid for a large data sample \cite{Wilks:1938fk}, it is not clear if the smoothed low-resolution sky map can qualify as a large data sample. In this paper we relax that assumption and calculate the confidence intervals exactly by estimating the distribution of the likelihood ratios using Monte-Carlo simulations. We use the ordering principle of Feldman and Cousins \cite{Feldman:1997qc} to estimate the confidence intervals. Let us briefly summarize the method in general and describe the usage of it for our case.

In general, the method works as follows. For a fixed point $P$ in the parameter space one constructs many simulations $S_i$ with the same values of the parameters $P$. For each simulation the best fit point $P_{best}$ in the parameter space is found and then the likelihood ratio calculated
\begin{equation}
R=\frac{\mathcal{L}(S_i|P)}{\mathcal{L}(S_i|P_{\rm best})}\,.
\end{equation}
Equivalently, one can calculate the difference in the logarithms of likelihood
\begin{equation}\label{likelihood_ratio_def}
\Delta\mathcal{L}=-2\ln\mathcal{L}(S_i|P)-(-2\ln\mathcal{L}(S_i|P_{\rm best}))\,.
\end{equation}
Then for a confidence level $\alpha$ one calculates a single number $\Delta\mathcal{L}_c$ such that $\alpha$ of the simulated experiments have $\Delta\mathcal{L}<\Delta\mathcal{L}_c$. For the real data $D$ the likelihood ratio is calculated for these values of the parameters $P$
\begin{equation}\label{likelihood_ratio_data}
\Delta\mathcal{L}_D=-2\ln\mathcal{L}(D|P)-(-2\ln\mathcal{L}(D|P_{\rm best}))
\end{equation}
and the point $P$ is accepted with confidence level $\alpha$ if $\Delta\mathcal{L}_D<\Delta\mathcal{L}_c$.

For our analysis we have four new parameters $L$, $\phi$, $\theta$, $\psi$, in addition to the standard cosmological parameters. The topology analysis is sensitive to the low-$l$ modes only while the standard cosmological parameters are determined from the whole range of $l$. Also, the standard cosmological parameters affect the diagonal elements of $M_{lml^\prime m^\prime}$ only, while the effects of topology are mainly in off-diagonal elements, as discussed in the previous section. This means that there is weak degeneracy between the standard parameters and the newly introduced ones. For this reason and computational cost, we fix the standard parameters to their best fit values as found from seven year WMAP data \cite{Komatsu:2010fb} and vary the new parameters only.\footnote{The values of the cosmological parameters that we use are $100\Omega_bh^2=2.227$, $\Omega_ch^2=0.1116$, $\Omega_{\Lambda}=0.729$, $n_s=0.966$, $\tau=0.085$, $\Delta_R^2(0.002\,\text{Mpc}^{-1})=2.42\times10^{-9}$.}

We would need to simulate many sky maps for all of the different possible values of the four new parameters, however, all of the different directions are equivalent for simulations. Namely, we expect the same distribution of likelihood ratios (\ref{likelihood_ratio_def}) for a fixed size $L$ but different directions. For a fixed topology and a fixed size $L$ we simulate $500$ sky maps for one fixed direction to find $\Delta\mathcal{L}_c$ for that size and direction, and use the same $\Delta\mathcal{L}_c$ for all of the directions with the same size.

For the real data we would need to find the likelihood ratio (\ref{likelihood_ratio_data}) for each point of the parameter space to decide if that point is accepted or not at a given confidence level. However, we are interested in the confidence intervals for the size $L$ only, therefore for each size $L$ we maximize the likelhood ratio over the angles and use that value to determine if the size $L$ must be accepted or not. This means that a given size $L$ is accepted at a given confidence level $\alpha$ if and only if there exists at least one point in the parameter space $(L, \phi, \theta, \psi)$ with that value of $L$ accepted at confidence level $\alpha$.

The simulations of finite sky maps are performed by first diagonalizing the covariance matrices (\ref{M_torus}) in harmonic space, then generating random gaussian samples from the diagonal matrices, then switching back to the original basis, and finally Fourier transforming into real space using the HEALPIX package  \cite{Gorski:2004by}.

\section{Results}\label{results_sec}

 We first do the analysis using the seven and nine year WMAP temperature maps. On large scales the seven year data already has negligible instrumental noise, so we do not expect a significant improvement in the results from nine year temperature data.
The best fit points for all the different maps are summarized in Table \ref{best_fit_0}. The errors on $L/L_0$ are simply determined by our scan step of $0.1$, while the errors on the Euler angles are found by the MINUIT package during the second fine scan. The $\mathcal{M}_2$ topology has one finite dimension (the $z$ axis) and two infinite dimensions, which means that it has the continuous symmetry of any rotations around the $z$ axis. For this reason, the final rotation by Euler angle $\psi$ does not change the topology, so the likelihood function does not depend on the angle. The Euler angles can be converted into the direction of the $z$ axis of the topologies in galactic coordinates by the simple prescription, $b=\frac{\pi}{2}-\theta$, and $l=\phi-\frac{\pi}{2}$.

\begin{figure}
\centering
\includegraphics[width=5cm]{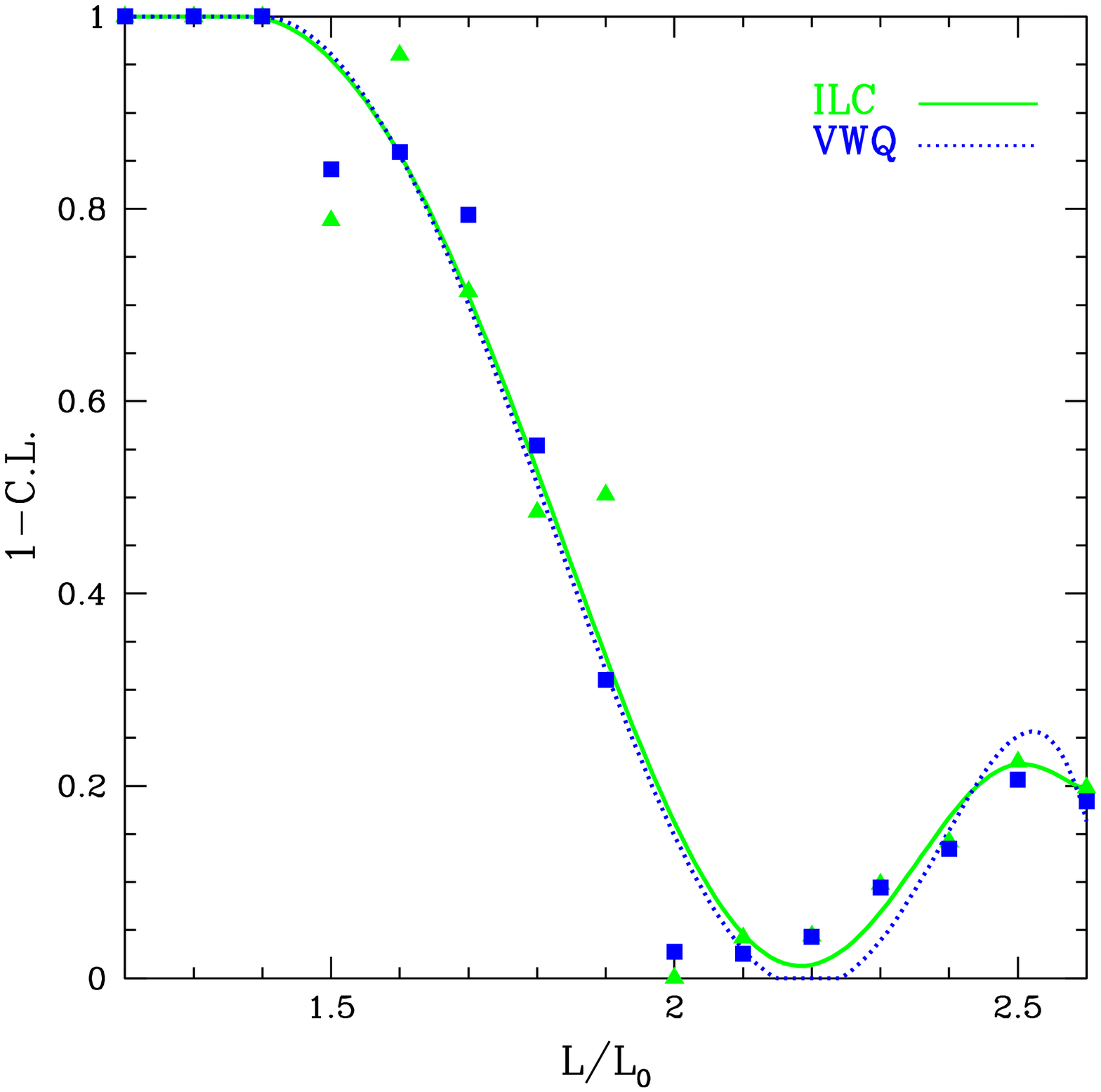}
\includegraphics[width=5cm]{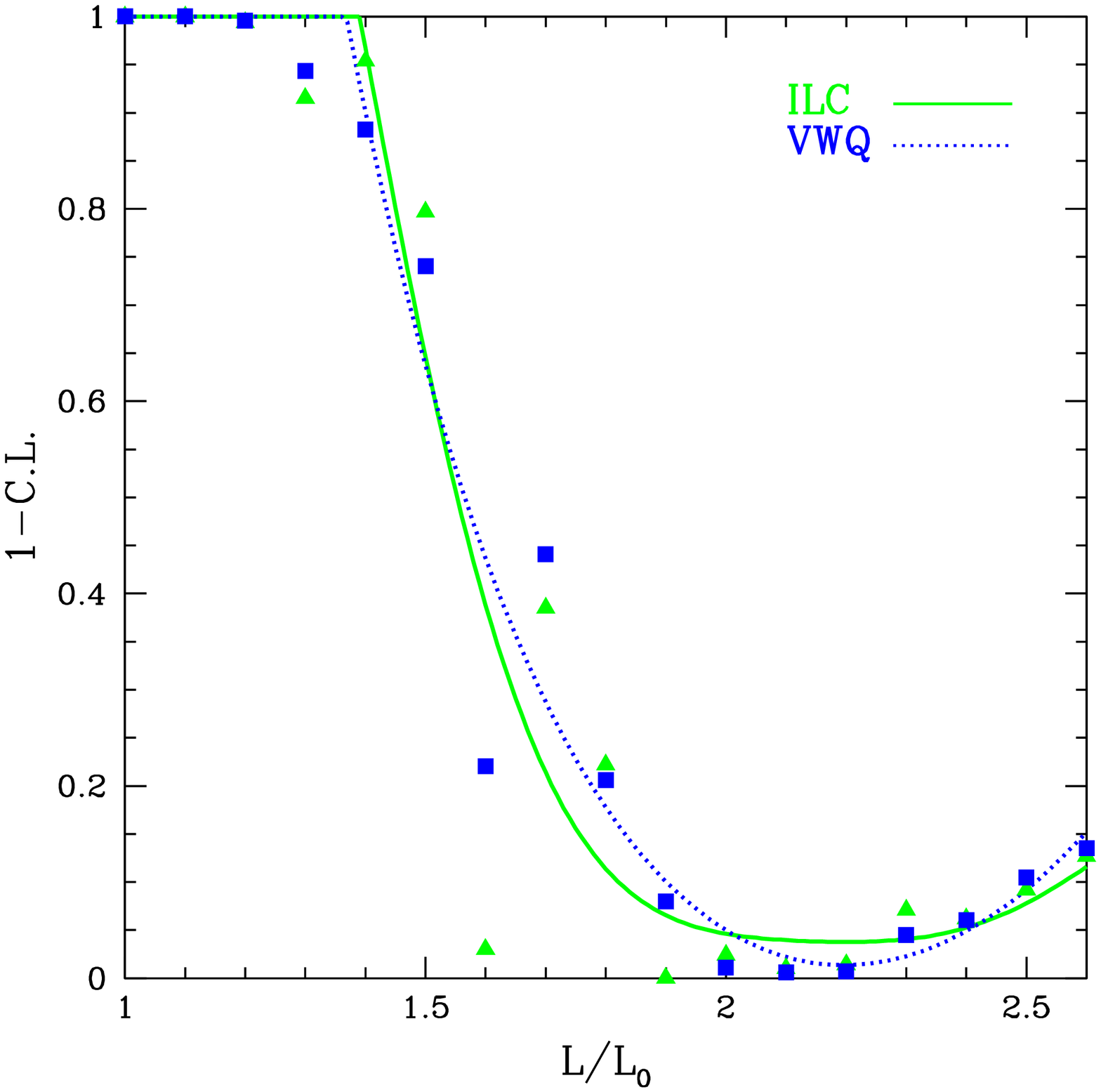}
\includegraphics[width=5cm]{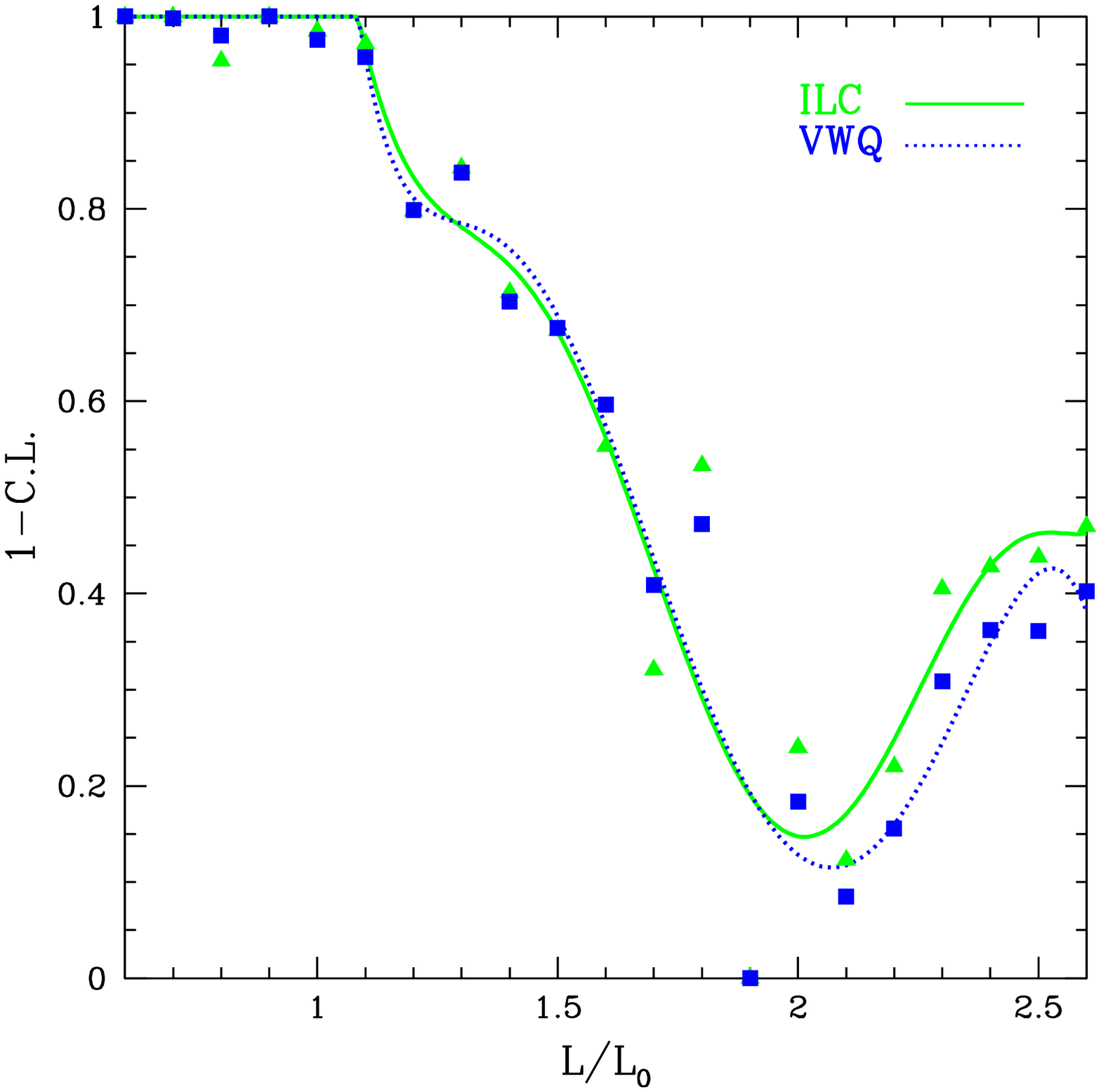}
\includegraphics[width=5cm]{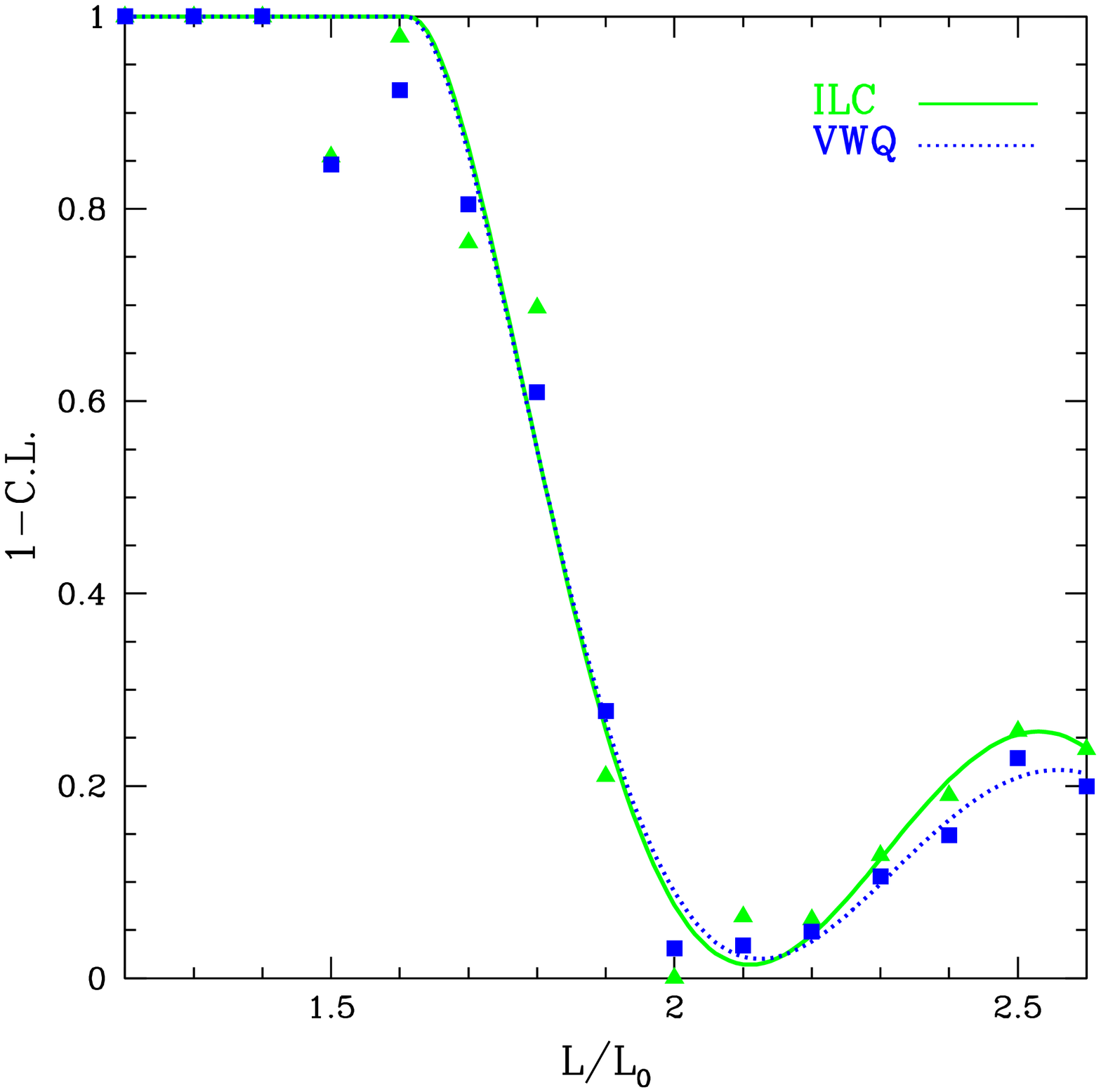}
\includegraphics[width=5cm]{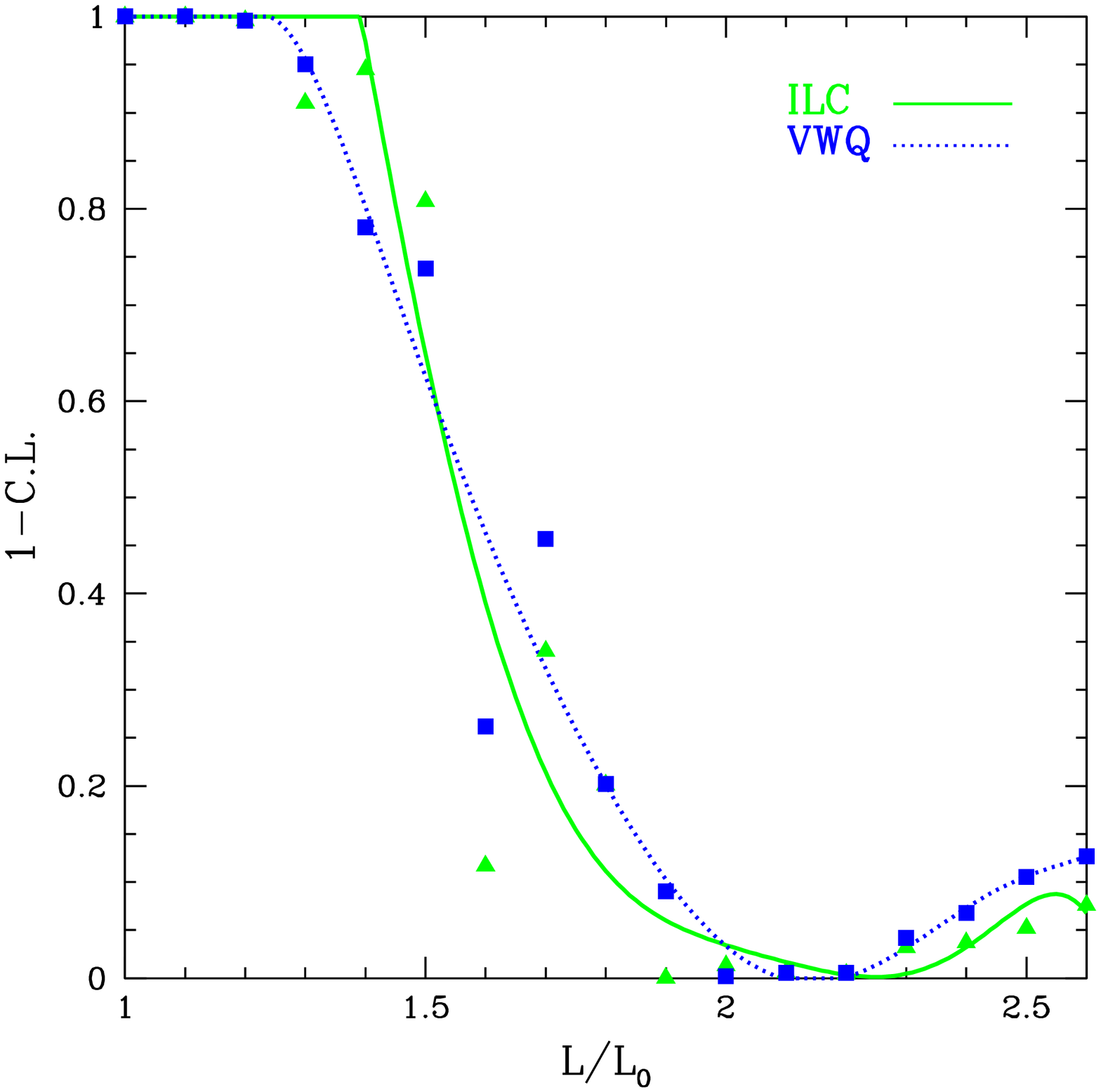}
\includegraphics[width=5cm]{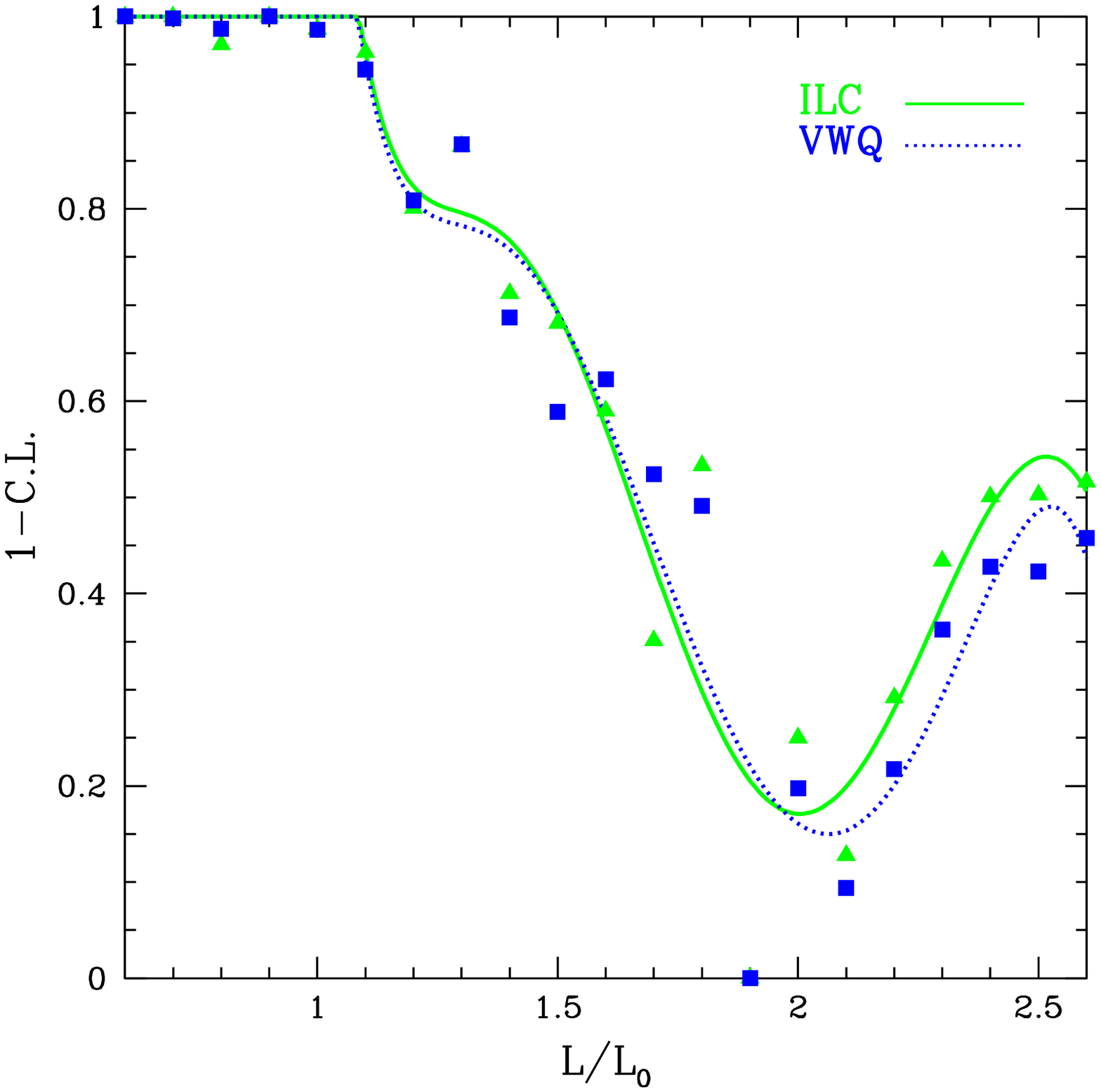}
\caption{\label{combined_fig} Plots of exclusion confidence level as a function of $L/L_0$ obtained from seven year and nine year WMAP temperature data only. The upper row corresponds to WMAP7, the lower row to WMAP9. The three plots correspond to $\mathcal{M}_0$ (left), $\mathcal{M}_1$ (middle), and $\mathcal{M}_2$ (right). The green triangular points correspond to the ILC map, the blue square points correspond to the combined data from V, W, and Q maps.
}
\end{figure}

\begin{figure}
\centering
\includegraphics[width=7.8cm]{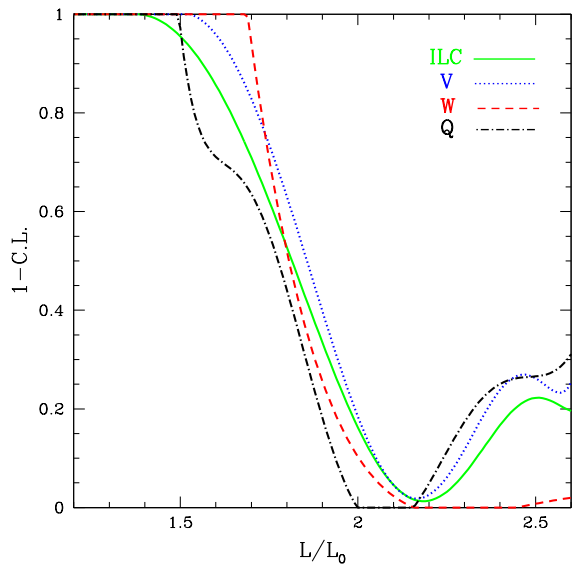}
\caption{\label{all_inf0_fig} Plots of exclusion confidence level as a function of $L/L_0$ obtained from seven year WMAP temperature data only for all of the maps analyzed. The  plot corresponds to the $\mathcal{M}_0$ topology.
}
\end{figure}

In order to obtain confidence intervals for the size $L$ we combine the results from V, W, and Q bands, and compare to the ILC results. Since we analyzed sizes with step $0.1L_0$ we fit a smooth curve to our data points to obtain more accurate results. The rejection confidence levels as a function of the size of the topology are shown in Fig. \ref{combined_fig} for the seven year and nine year WMAP temperature data. The smooth lines are a result of the fit to a $10$ degree polynomial. It is hard to estimate the errors associated with numerics and our discrete sampling, but the distance between our points and the smooth curve is a reasonable estimate of our error bars. For comparison, we show the results from separate maps in Fig. \ref{all_inf0_fig} for WMAP seven year temperature data for the $\mathcal{M}_0$ topology. The results from different maps agree reasonably well, however, after combining the V, W, and Q results the agreement with the ILC becomes much better. The $1\sigma$ and $2\sigma$ lower bounds for the size of the topologies are given in Table \ref{temperature_bounds}. The change of results from seven year to nine year data is very small, as expected.

\begin{table}
\renewcommand{\arraystretch}{1.5}
\setlength{\arraycolsep}{5pt}
\begin{eqnarray*}
\begin{array}{c|c|ccccc}
\text{Topology} &\text{Map} & \Delta\mathcal{L} & L/L_0 & \phi & \theta & \psi \\
\hline
\mathcal{M}_0  & \text{ILC (7)} & 18.89 & 2.0\pm0.05 & 2.328\pm0.036 & 2.512\pm0.012 & 0.379\pm0.033 \\
&\text{ILC (9)} & 19.45 & 2.0\pm0.05 & 2.330\pm0.035 & 2.512\pm0.012 & 0.380\pm0.033 \\
\hline
 \mathcal{M}_1&\text{ILC (7)} & 19.30 & 1.9\pm0.05 & 0.356\pm0.023 & 0.932\pm0.024 & 1.061\pm0.020 \\
&\text{ILC (9)} & 18.46 & 1.9\pm0.05 & 0.357\pm0.023 & 0.928\pm0.022 & 1.061\pm0.020 \\
\hline
\mathcal{M}_2&\text{ILC (7)} & 16.26 & 1.9\pm0.05 & 1.705\pm0.016 & 2.166\pm0.016 \\
&\text{ILC (9)} & 16.62 & 1.9\pm0.05 & 1.704\pm0.016 & 2.166\pm0.016 \\
%\text{V (7)} & 19.52 & 2.0\pm0.5 & 2.328\pm0.034 & 2.513\pm0.012 & 0.380\pm0.032 \\
%\text{V (9)} & 20.08 & 2.0\pm0.5 & 2.331\pm0.033 & 2.513\pm0.012 & 0.380\pm0.031 \\
%\text{W (7)} & 14.90 & 2.1\pm0.5 & 1.84\pm0.16 & 3.047\pm0.032 & 0.18\pm0.28 \\
%\text{W (9)} & 14.81 & 2.1\pm0.5 & 2.821\pm0.027 & 2.338\pm0.024 & 0.444\pm0.025 \\
%\text{Q (7)} & 19.99 & 2.0\pm0.5 & 2.089\pm0.047 & 2.730\pm0.014 & 0.348\pm0.043 \\
%\text{Q (9)} & 20.24 & 2.0\pm0.5 & 2.331\pm0.034 & 2.512\pm0.011 & 0.377\pm0.032 \\
\end{array}
\end{eqnarray*}
\caption{Best fit points and improvements in likelihood $\Delta\mathcal{L}$ compared to an infinite universe for the topology $\mathcal{M}_0, \mathcal{M}_1$ and $\mathcal{M}_2$ from seven year and nine year WMAP ILC temperature data. Note that for topology  $\mathcal{M}_2$ the likelihood does not depend on the Euler angle $\psi$.}  
\label{best_fit_0}
\end{table}

\begin{figure}
\centering
\includegraphics[width=7.8cm]{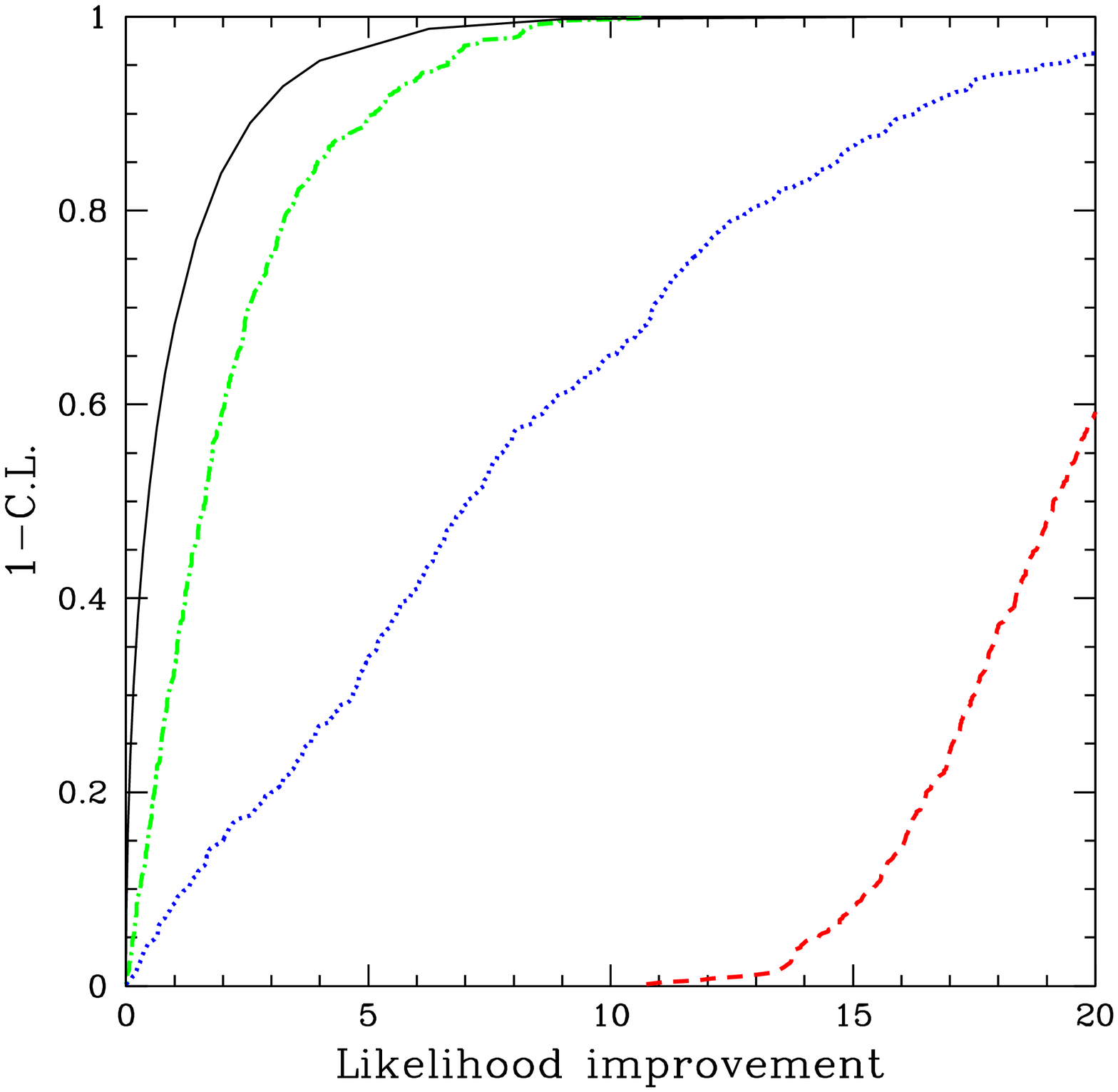}
\caption{\label{fc_data_fig} Exclusion confidence levels as a function of the improvement in likelihood $\Delta\mathcal{L}$. The black solid curve corresponds to the Gaussian approximation, the red dashed curve is found from simulations for an infinite universe, the blue dotted curve is from simulations for the $\mathcal{M}_0$ topology with size $L/L_0=2.0$, the green dash-dotted curve is from simulations for the $\mathcal{M}_0$ topology with size $L/L_0=1.0$.
}
\end{figure}

Let us now compare our results with the previous results in \cite{Aslanyan:2011zp} where seven year ILC temperature map was analyzed. The best fit point for the topology $\mathcal{M}_1$ is in close agreement with that found previously. For the other two topologies we find points that fit the data slightly better than those found previously. Note that the only difference in the analysis up to this point is a slightly better scan strategy over the angles. Therefore we find no contradiction to the previously found improvements in likelihood. However, after extracting the acceptance confidence limits for the size $L$ of the topologies from the likelihood improvements, we obtain confidence intervals that are larger than the ones found before. In particular, the infinite universe is accepted at about $1\sigma$ confidence level for all three topologies. These differences arise because of a completely different method of statistical data analysis used in this paper. The previous analysis~\cite{Aslanyan:2011zp} used the maximum likelihood method  which  relied on the assumption that the likelihood function is nearly gaussian in the size. In Fig. \ref{fc_data_fig} we plot the exclusion confidence level as a function of the improvement in likelihood (\ref{likelihood_ratio_def}) found from the Feldman-Cousins method compared to the theoretical approximation of a Gaussian distribution. As we can see, for small sizes ($L/L_0=1.0$ for the green dash-dotted curve) the Gaussian approximation works better than for larger sizes. For the best fit size of $L/L_0=2.0$ and for an infinite universe the curves from simulations are very far from the Gaussian curve. This is the reason why simulations are essential in order to find accurate confidence intervals.

\begin{table}
\renewcommand{\arraystretch}{1.5}
\setlength{\arraycolsep}{5pt}
\begin{eqnarray*}
\begin{array}{c|cccccc}
\text{Map} & \mathcal{M}_0(68.3 \%) & \mathcal{M}_0(95.5 \%) & \mathcal{M}_1(68.3\%) & \mathcal{M}_1(95.5\%) & \mathcal{M}_2(68.3\%) & \mathcal{M}_2(95.5\%)\\
\hline
\text{ILC (7)} & 1.71 & 1.50 & 1.49 & 1.40 & 1.49 & 1.11 \\
\text{VWQ (7)} & 1.71 & 1.50 & 1.48 & 1.38 & 1.50 & 1.10 \\
\text{ILC (9)} & 1.76 & 1.66 & 1.49 & 1.41 & 1.51 & 1.10 \\
\text{VWQ (9)} & 1.76 & 1.66 & 1.47 & 1.30 & 1.51 & 1.10 \\

%\text{ILC (7)} & 1.71 & 1.42 & 1.51 & 1.25 & 1.48 & 1.11 \\
%\text{ILC (9)} & 1.80 & 1.43 & 1.52 & 1.25 & 1.49 & 1.10 \\
%\text{V (7)} & 1.76 & 1.43 & 1.51 & 1.25 & 1.51 & 1.11 \\
%\text{V (9)} & 1.82 & 1.44 & 1.51 & 1.25 & 1.53 & 1.11 \\
%\text{W (7)} & 1.77 & 1.42 & 1.59 & 1.26 & 1.40 & 1.12 \\
%\text{W (9)} & 1.76 & 1.41 & 1.61 & 1.40 & 1.40 & 1.12 \\
%\text{Q (7)} & 1.59 & 1.51 & 1.44 & 1.31 & 1.52 & 1.00 \\
%\text{Q (9)} & 1.71 & 1.52 & 1.36 & 1.30 & 1.39 & 1.03 \\
\end{array}
\end{eqnarray*}
\caption{Lower bounds for $L/L_0$ at $1\sigma$ and $2\sigma$ confidence levels from seven year and nine year WMAP temperature data only.} 
\label{temperature_bounds}
\end{table}

\begin{figure}
\centering
\includegraphics[width=7.8cm]{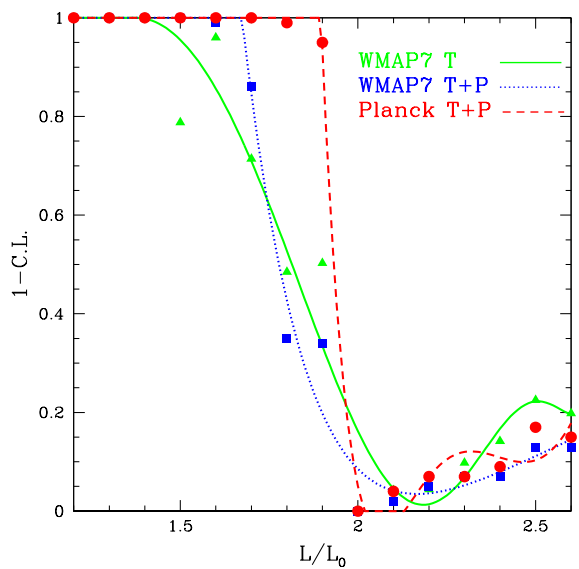}
\caption{\label{wmap_pol_fig} Plots of exclusion confidence level as a function of $L/L_0$ for the $\mathcal{M}_0$ topology obtained from seven year WMAP temperature data only, temperature and polarization data together, and forecast for infinite universes for Planck temperature and polarization data. The green triangular points correspond to WMAP7 temperature data only, the blue square points correspond to WMAP7 temperature and polarization data, and the red round points correspond to the forecast for Planck temperature and polarization data.
}
\end{figure}

Now we proceed to the results using the polarization data together with the temperature data. Since the WMAP polarization data is very noisy we do not expect much improvement in the results when we add the polarization data. For this reason we only analyze the WMAP seven year ILC map for the $\mathcal{M}_0$ topology only. The best fit point we find using the temperature and polarization data together corresponds to $L/L_0=2.0\pm0.05$, $\phi=0.446\pm0.016$, $\theta=1.792\pm0.020$, $\psi=0.595\pm0.014$. The improvement in $-2\ln\mathcal{L}$ is $17.57$. Comparing to the results for the same map and topology but with temperature data only (Table \ref{best_fit_0}, first row) we see that the best fit size stays the same, however the direction is different. Also, the improvement in likelihood for the best fit case compared to the infinite case is less than with the temperature data only.

Fig. \ref{wmap_pol_fig} compares the exclusion confidence levels as a function of the size of the topology for the analysis with and without the polarization data included. As we can see, including the polarization data slightly improves the lower bounds compared to the temperature data only. Using the polarization data we obtain the lower bound on $L/L_0$ of $1.73$ (compared to $1.71$ from temperature only) at $68.3\%$ confidence level and $1.68$ (compared to $1.50$ from temperature only) at $95.5\%$ confidence level.

Note that both for the real data and for simulated infinite universes we always find the best fit point near the size $2L_0$, varying between $1.9L_0$ and $2.2L_0$ as seen from Fig.~3. 
One would naively expect that an infinite universe would have a best fit at $L\rightarrow\infty$, however, by adding more parameters we allow the possibility of a better fit to the random quantum fluctuations. From Fig. \ref{diagnostics_fig} we can see that the size $\sim 2L_0$ is where the effects of a non-trivial topology start getting very small, so by minimizing over the orientation one is able to find a better fit to the fluctuations. This is another reason why doing simulations is essential. As can be seen in Figs. \ref{combined_fig} and \ref{planck_pol_fig} the better fit near the size $2L_0$ does not imply detection since $L\rightarrow\infty$ stays well within the $1\sigma$ confidence range, and this is true for both the real data and simulated infinite universes. The topology analysis by the Planck collaboration \cite{planck_topology} also found best fit points near the size $2L_0$ for both the real data and simulated infinite universes.

\section{Forecast for Planck}\label{forecast_sec}

The temperature maps from the Planck satellite have been analyzed to place lower bounds on the sizes of the three flat topologies discussed here, among other candidates, in \cite{planck_topology}. By maximizing the likelihood function over the directions they placed $95\%$ lower bounds on $L/L_0$ of $1.66$, $1.42$, and $1.00$ for the $\mathcal{M}_0$, $\mathcal{M}_1$, and $\mathcal{M}_2$ topologies, respectively. Compared to our results from Table \ref{temperature_bounds} from the WMAP data we can see that Planck is doing slightly better for the $\mathcal{M}_0$ and $\mathcal{M}_1$ topologies, but we have a slightly better lower bound for the $\mathcal{M}_1$ topology than Planck ($1.1$ instead of $1.0$). The analysis methods are similar to the ones used in this paper, except the scan strategy over directions. They used $10,000$ randomly chosen directions for the likelihood analysis which may not be enough to find the true maximum of the likelihood function. As discussed in Section \ref{data_analysis_sec} the variation of likelihood with the direction is very rapid and a fine scan is needed in order to find the true maximum. In the current work we used a similar number of directions for our initial scan, but we followed it by a finer scan to find the true maximum.

\begin{figure}
\centering
\includegraphics[width=7.8cm]{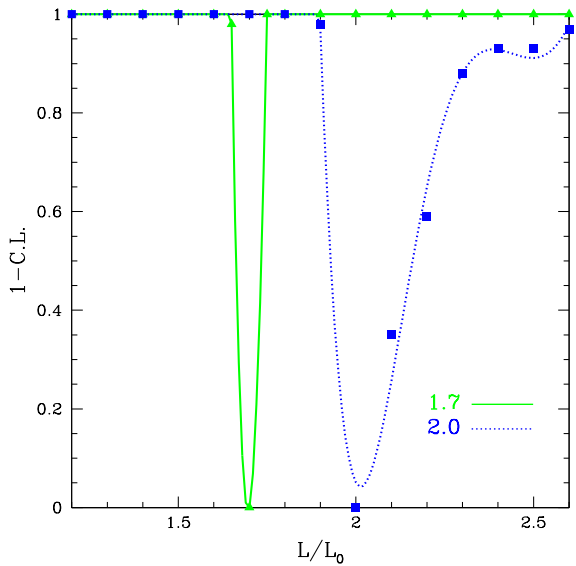}
\caption{\label{planck_pol_fig} Forecasts for exclusion confidence intervals for Planck for the $\mathcal{M}_0$ topology with sizes $1.7L_0$ and $2.0L_0$.
}
\end{figure}

The main reason why we introduced the formalism for including polarization data in the analysis is that the polarization data from the Planck satellite is expected to be released very soon, which will contain much less instrumental noise than the WMAP polarization maps. To get an idea of how the Planck data will improve the results we do a simple forecast for the $\mathcal{M}_0$ topology. We simulate $100$ infinite universes with the noise characteristics of Planck \cite{:2006uk}, find $\Delta\mathcal{L}_D$ from equation (\ref{likelihood_ratio_data}) for all of them, and use the median value to forecast the exclusion confidence levels for different sizes. The resulting confidence limits as a function of the size are shown in the red dashed curve of Fig. \ref{wmap_pol_fig}. We can see a significant improvement in lower bounds that Planck can place compared to WMAP. If the universe is infinite the expected lower bound on $L/L_0$ from Planck data is $1.92$ at $68.3\%$ confidence (vs. $1.73$ from WMAP) and $1.89$ at $95.5\%$ confidence (vs. $1.68$ from WMAP). The lower bounds obtainable from Planck polarization and temperature data are also better than the results from Planck temperature data alone ($1.89$ instead of $1.66$ at $2\sigma$ confidence level).

We also do similar simulations for the $\mathcal{M}_0$ topology with sizes $1.7L_0$ (slightly bigger than the lower bound set by Planck temperature data alone in \cite{planck_topology}) and $2.0L_0$ (bigger than any lower bound obtained so far from different methods). The results corresponding to these cases are shown Fig. \ref{planck_pol_fig}. Using the temperature and polarization data together the finite topology with size $1.7L_0$ can be detected at $3\sigma$ level at least (more simulations are needed to forecast detection at a higher level) with an error of less than $0.05L_0$. The finite topology with size $2.0L_0$ can be detected at about $2\sigma$ level with an error of $0.3L_0$. This proves that Planck polarization and temperature data together can put lower bounds on the size of an infinite universe comparable to the bounds that have been obtained from the alternative circles in the sky test, and can detect finite topologies with sizes greater than those accessible by that method.

\section{Discussion and Conclusions}\label{conclusions_sec}

In this paper we did a detailed analysis of the WMAP seven and nine year temperature maps for three flat spatial topologies of the universe. We used Monte-Carlo simulations and the Feldman-Cousins method to obtain confidence intervals for the size of the topologies analyzed. We further introduced the formalism for including the polarization data in the analysis and compared the results with and without the polarization maps for WMAP data for the $3$-torus topology $\mathcal{M}_0$. We then did a simple forecast for the Planck data.

One of our main conclusions is that the likelihood function is not Gaussian in the size of the topologies considered and simulations are essential for obtaining accurate confidence intervals. Although we find very similar improvements in likelihood as a previous analysis of the same topologies \cite{Aslanyan:2011zp}, using simulations changes the confidence levels for the sizes of the topologies significantly.

Using the temperature maps alone from WMAP seven year and nine year data we found lower bounds for the size of the topologies analyzed. For the topology $\mathcal{M}_0=\mathbb{T}^3$ the lower bound on the size $L/L_0$ at $95.5\%$ confidence level is about $1.5$, for the $\mathcal{M}_1=\mathbb{T}^2\times\mathbb{R}^1$ topology we find a lower bound of about $1.4$, and for the $\mathcal{M}_2=S^1\times\mathbb{R}^2$ topology the lower bound is about $1.1$. The infinite universe is excluded at less than $1\sigma$ confidence level for all three topologies. We analyzed the V, W, and $Q$ frequency bands in addition to the ILC map to make sure that the results are similar. We found no significant improvement from nine year WMAP temperature data compared to the seven year data. Our results agree very well with the corresponding results from Planck temperature data \cite{planck_topology}. The measurement of temperature maps on large scales is not improved by Planck compared to WMAP (except for better foreground removal), so it is expected to get similar results from these data sets. Therefore, our results from temperature maps confirm independently the results from Planck.

Including polarization data from WMAP slightly improves the results obtained from the temperature data alone. However, a simple forecast shows that the improvement will be significant for Planck results (the $95.5\%$ confidence lower bound on the size $L/L_0$ for the $\mathcal{M}_0$ topology will improve to about $1.9$ from $1.7$ for an infinite universe, and a finite universe of size $2.0L_0$ can be detected at $95.5\%$ confidence level), so the techniques developed in this paper will be very useful for a detailed analysis of the Planck temperature and polarization maps. Also, the correlation between the temperature and polarization maps can serve as an independent check of the results. If similar signatures are seen in both temperature and polarization maps then it will be very unlikely that the effect is a result of some systematics in the temperature data or not completely removed foreground.

\acknowledgments

The authors would like to thank the LAMBDA team for making the seven year and nine year WMAP data and the likelihood calculation software available online, and Antony Lewis and Anthony Challinor for making the CAMB software available online. We thank Layne Price for helpful comments on the manuscript.

The authors wish to acknowledge the contribution of the NeSI high-performance computing facilities and the staff at the Centre for eResearch at the University of Auckland. New Zealand's national facilities are provided by the New Zealand eScience Infrastructure (NeSI) and funded jointly by NeSI's collaborator institutions and through the Ministry of Business, Innovation and Employment's Infrastructure programme. URL {\tt http:/\!/www.nesi.org.nz}. This work was also supported in part by the U.S. Department of Energy through DOE Grant No. DE-FG02-90ER40546.

\bibliography{citations}

\providecommand{\href}[2]{#2}\begingroup\raggedright\begin{thebibliography}{10}

\bibitem{deOlivieraCosta:1994eb}
A.~de~Oliviera~Costa and G.~F. Smoot, {\it {Constraints on the topology of the
  universe from the 2- year COBE data}},  {\em Astrophys. J.} {\bf 448} (1995)
  477, [\href{http://xxx.lanl.gov/abs/astro-ph/9412003}{{\tt
  astro-ph/9412003}}].

\bibitem{deOliveiraCosta:1995td}
A.~de~Oliveira-Costa, G.~F. Smoot, and A.~A. Starobinsky, {\it {Can the lack of
  symmetry in the COBE/DMR maps constrain the topology of the universe?}},
  {\em Astrophys. J.} {\bf 468} (1996) 457,
  [\href{http://xxx.lanl.gov/abs/astro-ph/9510109}{{\tt astro-ph/9510109}}].

\bibitem{Phillips:2004nc}
N.~G. Phillips and A.~Kogut, {\it {Constraints On The Topology Of The Universe
  From The WMAP First-Year Sky Maps}},  {\em Astrophys. J.} {\bf 645} (2006)
  820--825, [\href{http://xxx.lanl.gov/abs/astro-ph/0404400}{{\tt
  astro-ph/0404400}}].

\bibitem{Kunz:2005wh}
M.~Kunz {\em et.~al.}, {\it {Constraining topology in harmonic space}},  {\em
  Phys. Rev.} {\bf D73} (2006) 023511,
  [\href{http://xxx.lanl.gov/abs/astro-ph/0510164}{{\tt astro-ph/0510164}}].

\bibitem{Aurich:2007yx}
R.~Aurich, H.~S. Janzer, S.~Lustig, and F.~Steiner, {\it {Do we Live in a
  'Small Universe'?}},  {\em Class. Quant. Grav.} {\bf 25} (2008) 125006,
  [\href{http://xxx.lanl.gov/abs/0708.1420}{{\tt arXiv:0708.1420}}].

\bibitem{Aslanyan:2011zp}
G.~Aslanyan and A.~V. Manohar, {\it {The Topology and Size of the Universe from
  the Cosmic Microwave Background}},  {\em JCAP} {\bf 1206} (2012) 003,
  [\href{http://xxx.lanl.gov/abs/1104.0015}{{\tt arXiv:1104.0015}}].

\bibitem{Roukema:2010mw}
B.~F. Roukema, {\it {Which FLRW comoving 3-manifold is preferred
  observationally and theoretically?}},
  \href{http://xxx.lanl.gov/abs/1002.3528}{{\tt arXiv:1002.3528}}.

\bibitem{Luminet:2003dx}
J.~P. Luminet, J.~Weeks, A.~Riazuelo, R.~Lehoucq, and J.~P. Uzan, {\it
  {Dodecahedral space topology as an explanation for weak wide-angle
  temperature correlations in the cosmic microwave background}},  {\em Nature.}
  {\bf 425} (2003) 593, [\href{http://xxx.lanl.gov/abs/astro-ph/0310253}{{\tt
  astro-ph/0310253}}].

\bibitem{Caillerie:2007gd}
S.~Caillerie {\em et.~al.}, {\it {A new analysis of Poincar\'e dodecahedral
  space model}},  {\em Astron. Astrophys.} {\bf 476} (2007), no.~2 691--696,
  [\href{http://xxx.lanl.gov/abs/0705.0217}{{\tt arXiv:0705.0217}}].

\bibitem{Lew:2008yz}
B.~S. Lew and B.~F. Roukema, {\it {A test of the Poincare dodecahedral space
  topology hypothesis with the WMAP CMB data}},  {\em Astron. Astrophys.} {\bf
  482} (2008), no.~3 747--753, [\href{http://xxx.lanl.gov/abs/0801.1358}{{\tt
  arXiv:0801.1358}}].

\bibitem{Aurich:2004fq}
R.~Aurich, S.~Lustig, and F.~Steiner, {\it {CMB Anisotropy of the Poincare
  Dodecahedron}},  {\em Class. Quant. Grav.} {\bf 22} (2005) 2061--2083,
  [\href{http://xxx.lanl.gov/abs/astro-ph/0412569}{{\tt astro-ph/0412569}}].

\bibitem{Roukema:2011xm}
B.~F. Roukema and T.~A. Kazimierczak, {\it {The size of the Universe according
  to the Poincare dodecahedral space hypothesis}},  {\em Astron.Astrophys.}
  {\bf 533} (2011) A11, [\href{http://xxx.lanl.gov/abs/1106.0727}{{\tt
  arXiv:1106.0727}}].

\bibitem{2011MNRAS.tmp..137B}
P.~{Bielewicz} and A.~J. {Banday}, {\it {Constraints on the topology of the
  Universe derived from the 7-yr WMAP data}},  {\em MNRAS} {\bf 412} (2011)
  2104, [\href{http://xxx.lanl.gov/abs/1012.3549}{{\tt arXiv:1012.3549}}].

\bibitem{Aurich:2013}
R.~Aurich and S.~Lustig, {\it {A search for cosmic topology in the final WMAP
  data}},  \href{http://xxx.lanl.gov/abs/1303.4226}{{\tt arXiv:1303.4226}}.

\bibitem{planck_topology}
{Planck Collaboration}, {\it {Planck 2013 results. XXVI. Background geometry
  and topology of the Universe}},
  \href{http://xxx.lanl.gov/abs/1303.5086}{{\tt arXiv:1303.5086}}.

\bibitem{Bielewicz:2011jz}
P.~Bielewicz, A.~Banday, and K.~Gorski, {\it {Constraining the topology of the
  Universe using the polarised CMB maps}},  {\em MNRAS} {\bf 421(2)} (2012)
  1064--1072, [\href{http://xxx.lanl.gov/abs/1111.6046}{{\tt
  arXiv:1111.6046}}].

\bibitem{Feldman:1997qc}
G.~J. Feldman and R.~D. Cousins, {\it {A Unified Approach to the Classical
  Statistical Analysis of Small Signals}},  {\em Phys. Rev.} {\bf D57} (1998)
  3873--3889, [\href{http://xxx.lanl.gov/abs/physics/9711021}{{\tt
  physics/9711021}}].

\bibitem{Page:2006hz}
{\bf WMAP Collaboration} Collaboration, L.~Page {\em et.~al.}, {\it {Three year
  Wilkinson Microwave Anisotropy Probe (WMAP) observations: polarization
  analysis}},  {\em Astrophys.J.Suppl.} {\bf 170} (2007) 335,
  [\href{http://xxx.lanl.gov/abs/astro-ph/0603450}{{\tt astro-ph/0603450}}].

\bibitem{2012arXiv1212.5226H}
G.~{Hinshaw}, D.~{Larson}, E.~{Komatsu}, D.~N. {Spergel}, C.~L. {Bennett},
  J.~{Dunkley}, M.~R. {Nolta}, M.~{Halpern}, R.~S. {Hill}, N.~{Odegard},
  L.~{Page}, K.~M. {Smith}, J.~L. {Weiland}, B.~{Gold}, N.~{Jarosik},
  A.~{Kogut}, M.~{Limon}, S.~S. {Meyer}, G.~S. {Tucker}, E.~{Wollack}, and
  E.~L. {Wright}, {\it {Nine-Year Wilkinson Microwave Anisotropy Probe (WMAP)
  Observations: Cosmological Parameter Results}},  {\em ArXiv e-prints} (Dec.,
  2012) [\href{http://xxx.lanl.gov/abs/1212.5226}{{\tt arXiv:1212.5226}}].

\bibitem{Lewis:1999bs}
A.~Lewis, A.~Challinor, and A.~Lasenby, {\it Efficient computation of {CMB}
  anisotropies in closed {FRW} models},  {\em Astrophys. J.} {\bf 538} (2000)
  473--476, [\href{http://xxx.lanl.gov/abs/astro-ph/9911177}{{\tt
  astro-ph/9911177}}].

\bibitem{Jarosik:2010iu}
N.~Jarosik {\em et.~al.}, {\it {Seven-Year Wilkinson Microwave Anisotropy Probe
  (WMAP) Observations: Sky Maps, Systematic Errors, and Basic Results}},  {\em
  Astrophys. J. Suppl.} {\bf 192} (2011) 14,
  [\href{http://xxx.lanl.gov/abs/1001.4744}{{\tt arXiv:1001.4744}}].

\bibitem{Larson:2010gs}
D.~Larson {\em et.~al.}, {\it {Seven-Year Wilkinson Microwave Anisotropy Probe
  (WMAP) Observations: Power Spectra and WMAP-Derived Parameters}},  {\em
  Astrophys. J. Suppl.} {\bf 192} (2011) 16,
  [\href{http://xxx.lanl.gov/abs/1001.4635}{{\tt arXiv:1001.4635}}].

\bibitem{Komatsu:2010fb}
E.~Komatsu {\em et.~al.}, {\it {Seven-Year Wilkinson Microwave Anisotropy Probe
  (WMAP) Observations: Cosmological Interpretation}},  {\em Astrophys. J.
  Suppl.} {\bf 192} (2011) 18, [\href{http://xxx.lanl.gov/abs/1001.4538}{{\tt
  arXiv:1001.4538}}].

\bibitem{Gorski:2004by}
K.~Gorski, E.~Hivon, A.~Banday, B.~Wandelt, F.~Hansen, {\em et.~al.}, {\it
  {HEALPix - A Framework for high resolution discretization, and fast analysis
  of data distributed on the sphere}},  {\em Astrophys.J.} {\bf 622} (2005)
  759--771, [\href{http://xxx.lanl.gov/abs/astro-ph/0409513}{{\tt
  astro-ph/0409513}}].

\bibitem{Hinshaw:2006ia}
{\bf WMAP} Collaboration, G.~Hinshaw {\em et.~al.}, {\it {Three-year Wilkinson
  Microwave Anisotropy Probe (WMAP) observations: temperature analysis}},  {\em
  Astrophys.J.Suppl.} {\bf 170} (2007) 288,
  [\href{http://xxx.lanl.gov/abs/astro-ph/0603451}{{\tt astro-ph/0603451}}].

\bibitem{Aslanyan:2013zs}
G.~Aslanyan, A.~V. Manohar, and A.~P. Yadav, {\it {Limits on Semiclassical
  Fluctuations in the Primordial Universe}},  {\em JCAP} {\bf 1302} (2013) 040,
  [\href{http://xxx.lanl.gov/abs/1301.5641}{{\tt arXiv:1301.5641}}].

\bibitem{minuit}
``http://seal.web.cern.ch/seal/work-packages/mathlibs/minuit.''

\bibitem{Wilks:1938fk}
S.~Wilks, {\it The large-sample distribution of the likelihood ratio for
  testing composite hyptheses},  {\em Ann. Math. Stat.} {\bf 9} (1938), no.~1
  60--62.

\bibitem{:2006uk}
{\bf Planck} Collaboration, {Planck Collaboration}, {\it {Planck: The
  scientific programme}},  \href{http://xxx.lanl.gov/abs/astro-ph/0604069}{{\tt
  astro-ph/0604069}}.

\end{thebibliography}\endgroup

\end{document}